\documentclass[journal,subeqn,caption]{IEEEtran}

\usepackage[pdftex]{graphicx}
\usepackage{amssymb,amsfonts,amsmath,mathrsfs}
\usepackage{bm}
\usepackage{color}
\usepackage{url}
\usepackage{cite}

\newcommand{\pW}{\mathbb{P}_{\Omega}}
\newcommand{\pw}{\mathbb{P}_{\omega}}

\newcommand{\cE}{{\cal E}}
\newcommand{\cV}{{\cal V}}

\newcommand{\fA}{{\forall}}

\newcommand{\generationsetpoint}{\bar{p}}

\allowdisplaybreaks

\newcommand{\edit}[1]{{\color{black}{#1}}}

\newcommand{\avz}[1]{{\color{black}{#1}}}

\begin{document}
\title{An Uncertainty Management Framework for Integrated Gas-Electric Energy Systems}

\author{Line Roald,
        Kaarthik Sundar,
        Anatoly Zlotnik,
        Sidhant Misra, and
        G\"oran Andersson %
\thanks{L. Roald is with the Department
of Electrical and Computer Engineering, University of Wisconsin-Madison, Madison, WI, 53706 USA, e-mail: roald@wisc.edu. 
K. Sundar, A. Zlotnik, and S. Misra are with Los Alamos National Laboratory, Los Alamos, NM, 87545 USA, e-mail: \{kaarthik $\mid$ azlotnik $\mid$ sidhant \}@lanl.gov. G\"oran Andersson is Professor Emeritus at ETH Zurich, Zurich,
Switzerland, e-mail: andersson@eeh.ee.ethz.ch.}
\thanks{The first author acknowledges funding from the Center for Non-Linear Studies (CNLS) at  Los Alamos National Laboratory and U.S. Department of Energy, Office of Science, Office of Advanced Scientific Computing Research, Applied Mathematics program under Contract Number DE-AC02-06CH11347.  The authors are grateful to Sandra Jenkins and Alireza Ghassemian of the U.S. Department of Energy Office of Electricity for supporting this study.  Work at Los Alamos National Laboratory was conducted under the auspices of the National Nuclear Security Administration of the U.S. Department of Energy under Contract
No. 89233218CNA000001.  This manuscript is approved for unlimited distribution with report number LA-UR-20-20818.}}

\maketitle

\begin{abstract}
\edit{In many parts of the world, electric power systems have seen a significant shift towards generation from renewable energy and natural gas. Because of their ability to flexibly adjust power generation in real time, gas-fired power plants are frequently seen as the perfect partner for variable renewable generation. However, this reliance on gas generation increases interdependence and propagates uncertainty between power grids and gas pipelines, and brings coordination and uncertainty management challenges. To address these issues, we propose an uncertainty management framework for uncertain, but bounded gas consumption by gas-fired power plants. The admissible ranges are computed based on a joint optimization problem for the combined gas and electricity networks, which involves chance-constrained scheduling for the electric grid and a novel robust optimization formulation for the natural gas network.  This formulation ensures feasibility of the integrated system with a high probability, while providing a tractable numerical formulation. A key advance with respect to existing methods is that our method is based on a physically accurate, validated model for transient gas pipeline flows. 
Our case study benchmarks our proposed formulation against methods that ignore how reserve activation impacts the fuel use of gas power plants, and only consider predetermined gas consumption. The results demonstrate the importance of considering uncertainty to avoid operating constraint violations and curtailment of gas to the generators.}

\end{abstract}

\IEEEpeerreviewmaketitle

\section{Introduction} \label{sec:intro}
\bstctlcite{}
Throughout the world, the share of electricity generated from renewable energy sources is rapidly increasing, concurrently with a shift towards using natural gas as the predominant fuel for traditional bulk generation \cite{IEA2019}.  These trends are both driven by near-term goals to decrease carbon emissions \cite{Bistline2014}, as well as long term plans to progress towards fully renewable electric energy systems \cite{DominguezConejoCarrion2015}. 
In the United States, technological changes that enable the exploitation of previously inaccessible natural gas resources, such as shale formations, have made this fuel abundant and inexpensive \cite{levitan14,costello13}.  Conveniently, natural gas-fired generators produce significantly lower carbon dioxide and particulate emissions relative to coal power plants, and are much easier to site, permit, and build than nuclear generating stations \cite{Bistline2015}.  This has placed natural gas as the ``transition fuel'' in the evolving generation mix, which is viewed as a bridge between the traditional centralized grid and a future sustainable energy system \cite{moniz2011future,logan2013natural,aguilera2019revisiting}. 

\avz{The advantages of gas-fired generators go beyond reduced cost and emissions, and include the ability to flexibly modulate power production in real time.} Single cycle plants can quickly go online and have flexible ramping capabilities that allow for fast modulation of their generation output. They are therefore often used as marginal resources that start and shut down multiple times a day. This capability is viewed as filling the near-term need to compensate for rapid changes in uncontrollable renewable generation \cite{BADAKHSHAN2019844,HittingerWhitacreApt2010}, but also leads to high-volume changes and short-term uncertainty in natural gas consumption \cite{chertkov15ae}. 
The characteristics of gas-fired power plants are hence very different from traditional loads on natural gas pipelines, such as local gas distribution companies, who hold firm delivery contracts for steady supply of natural gas \cite{rubio08}, or industrial customers, who have time-varying, but predictable gas withdrawals\cite{chertkov15ae, chertkov15hicss}. 
As gas-fired generation expands to account for nearly 50\% of installed electric generation capacity in the U.S., such differences become increasingly problematic, and make the power grid vulnerable to interruptions in the natural gas supply chain \cite{babula2014cold,pugh10,shahidehpour14}. These interruptions increasingly challenge the ability of power grid operators to meet generation demand, maintain operating reserves, and ensure power system reliability \cite{isone12}.

Addressing the risk of natural gas shortfall to electric generators is thus essential, %
and requires the development of more efficient and reliable decision support tools to manage the impact of more variable and uncertain operations in a secure, economical manner. %
\edit{These new tools must address three important challenges:}

\edit{First,} there is a need for \emph{increased coordination} and information exchange between gas pipelines and the power grid, particularly to improve transparency and jointly manage \avz{constraints of the integrated energy delivery network, which we refer to as the \emph{gas-electric system}.} %

\edit{Second, the tools must include models that reflect the \emph{physics of energy flow} in electric grids and natural gas pipelines, which each} have their distinct challenges. 
Because electric grids have minimal inherent storage, grid operators must balance generation with load at all times \cite{wood13}, which makes it challenging to secure the system against large outages or fluctuations in renewable energy. 
\edit{The fast system dynamics of electric grids enables power flow in these systems to be accurately represented using steady-state power flow models.}
Natural gas pipelines, on the other hand, have significant energy stored internally in the form of the transported gas, and can sustain unbalanced operations over hours or days \cite{carter03}. 
\edit{However, representing this behavior requires modeling of \emph{transient gas dynamics}, which present modeling and computation challenges.}

\edit{Third, because gas-fired generators are essential in balancing variations in renewable energy, there is a need for tools that enable \emph{uncertainty management} across the interface between gas and electricity networks. Specifically, power grid operators must know how they can control gas-fired generators to ensure balance in the electric grid, without adversely impacting the integrity and reliability of natural gas pipelines. Because renewable energy uncertainty occurs at the time-scale of intra-day operation, where natural gas systems are not in steady-state, it is crucial that the model accurately captures the impact of uncertainty on the transient gas dynamics.}

\avz{Previous studies, which we review extensively in the next section, have considered several of the above aspects.} However, to the best of our knowledge,
there are no existing methods that constructively address \emph{integrated optimization} of natural gas and electricity delivery considering \emph{transient} natural gas dynamics and \emph{uncertainty} from renewable energy. \edit{Combining these three aspects is essential for developing effective decision support tools. Unfortunately, the simultaneous consideration of joint optimization, uncertainty, and transient gas flows creates very challenging non-convex, large-scale problems.} %
In this study, we address this open challenge.

\edit{Specifically, the main contribution of our study is to propose \emph{the first method for operational scheduling of gas-electric energy systems that jointly considers integrated optimization, coordinated uncertainty management, and gas pipeline transients}. 
The core idea of our framework is simple, but powerful. %
Our method determines three natural gas consumption profiles corresponding to the scheduled, minimum, and maximum gas withdrawals for each gas-fired generator at each point in time. These three profiles are obtained from the solution of a joint optimization framework for the integrated gas-electric system.
In the power system, the gas withdrawal profiles represent the range of available flexibility from the gas-fired generators. By utilizing a chance-constrained optimization formulation for power system operations, we can guarantee (with a high probability) that these withdrawal profiles are consistent with the need for reserve activation in the electric grid.
In the natural gas pipelines, the gas withdrawal profiles represent acceptable uncertainty in consumption. Given the maximum and minimum profiles, we apply recent monotonicity results \cite{vuffray15cdc,misra2016monotone,misra20pieee} to formulate robust feasibility conditions for the transient natural gas flow dynamics. By including these robust feasibility conditions in the joint optimization, we guarantee that \emph{any} gas withdrawal profile between the maximum and minimum profiles will not violate the gas system constraints. \emph{Thus, by enforcing feasibility for only three gas withdrawal profiles, we are able to guarantee viability of our solutions for a wide range of practical operational scenarios.}
}

\avz{We wish to highlight that the formulation of a tractable intra-day optimization problem that simultaneously considers the impact of uncertainty both in the electric grid and gas pipeline transients is nontrivial and novel.
Crucially, the formulation leverages recent monotonicity results for transient gas pipeline flows, which enable a new approach to robust gas network optimization and a simple three-scenario interface for information exchange between gas pipeline and power grid operators.}

The remainder of the paper is organized as follows.  In Section \ref{sec:interactions} we provide a detailed review \edit{of related work from} the quickly evolving literature on gas-electric system coordination, \edit{with sections dedicated to each of the three challenges of coordination and integrated optimization, modeling of the transient natural gas dynamics, and uncertainty management}. Section \ref{sec:framework} summarizes \edit{the conclusions from our literature review} and provides an overview of the framework for modeling and coordination of electric grids and natural gas pipelines under uncertainty, based on \emph{uncertain, but bounded} gas withdrawals. 
Section \ref{sec:power} describes the modeling and optimization framework for multi-period, chance-constrained power system scheduling.
In Section \ref{sec:gas}, we provide an overview of the modeling and optimization of large-scale natural gas pipelines, before we describe the extension to robust optimization utilizing monotone systems properties.
Based on the formulations provided in the previous sections, Section \ref{sec:integration} details the coupling between the two systems, determined by the gas-fired generators and the energy and reserves they are scheduled to provide, and presents the mathematical model for joint scheduling and uncertainty management. 
Section \ref{sec:computation} outlines the framework we utilize to assess the performance of the method, as well as two benchmark formulations. The numerical results presented in Section \ref{sec:results} 
demonstrate the performance of the method, and show that accounting for uncertainty of the gas withdrawals is essential for reliable delivery of both gas and electricity.
Section \ref{sec:conclusion} summarizes and concludes.

\section{\edit{Review of Interactions between Electric Energy Systems and Natural Gas Pipelines}} \label{sec:interactions}
As the bulk electric power system increasingly relies on gas-fired generation, it becomes critical to study the interactions between these complex networked systems, and develop adequate modeling and optimization techniques to coordinate and manage the integrated gas-electric system.
In this section we review key issues relating these interacting infrastructure systems, as well as recent academic literature and industry studies. \edit{We divide our review into three parts corresponding to the three challenges identified in the introduction, namely A) optimization and coordination of the integrated system, B) modeling of transient gas flow dynamics, and C) uncertainty management. 
}

\subsection{\edit{Integrated Optimization and Coordination of Electric and Natural Gas Transmission Networks}}
\edit{Current interactions between natural gas and electric systems are primarily driven by the electricity demand from natural gas fired power plants.
Because gas-fired power plants do not store fuel onsite, gas is a ``just-in-time'' fuel that must be continuously available in operation. Fuel supply interruptions thus incur a risk of simultaneous outage of multiple gas-fired generators. This has led to growing concerns about inter-sector coordination \cite{isone12,black13,mitei14}, leading to regulatory action in the U.S. \cite{carter16}. 
The situation is particularly challenging because power plants often do not hold firm contracts for pipeline transportation, and are not guaranteed access to sufficient pipeline capacity to transport the gas needed for generation. Instead, gas-fired generators typically rely on shorter-term arrangements in the gas pipeline capacity release market. Thus, in periods when pipeline capacity is subscribed entirely by firm contract holders (e.g., during very cold weather), power plants are the first loads to be curtailed \cite{myles2017ensuring}. }

\subsubsection{\edit{Optimization of Electric Grid Operations}}
The fuel usage of these gas generators is determined primarily by the production schedules created in day-ahead electricity markets. In the U.S., these markets are cleared by independent (electric) system operators (ISOs) that solve unit commitment, reserve allocation, and economic dispatch problems based on optimal power flow (OPF) formulations \cite{litvinov10, pjm15, isone15}.  
The optimization based markets give rise to electricity prices that are consistent with the physical capacity of the power grid, and have enabled much more efficient utilization of generation and transmission assets \cite{wang2017deregulation,razeghi2017impact,lindsay2019power}. \avz{The paradigm in power grid operations is integration of market clearing, scheduling, and operations by standardized optimization-based decision support.}

\subsubsection{\edit{Optimization of Natural Gas Pipeline Operations}}
As the use of natural gas for electricity generation continues to increase,
it is important to increase responsiveness and efficiency of natural gas pipeline operations,
particularly as the development of new infrastructure slows \cite{feijoo2018future,oliver2015economies}.     
Historical load and price analysis in the U.S. natural gas  markets indicates that price spikes occur when load levels approach 75\% of firm contract capacity \cite{black13}, which is conventionally taken as the constraint capacity threshold \cite{koch15}. \avz{The prevailing paradigm of pipeline operations is the to conduct marketing, scheduling, and physical gas control operations as separate activities, often using labor intensive experience-based approaches developed by individual companies.}
One path to more responsive and efficient operation of gas pipelines 
is the integration of markets and physical control using optimization-based decision support systems analogous to those used in electricity markets.
\edit{In gas pipeline systems, flow is controlled using compressors and regulators that directly change pressures and flow rates, or by scheduling injection and consumption of gas. }
Optimization-based scheduling of pipeline system operations can 
provide location- and time-dependent prices for natural gas that account for pipeline flow physics, engineering limitations, and operational constraints \cite{zlotnik17psig,rudkevich17hicss,novitsky2019multilevel}. 
This would allow gas-fired generation asset managers to provide bids and offers that reflect their price-sensitive demand \cite{zhao2018shadow}.
Advancing pipeline operations methods will also increase reliability in supplying fuel to the electric power sector during highly loaded conditions \cite{zlotnik17tpwrs}.
\edit{This may become increasingly important in the future, when power-to-gas technology, which converts excess renewable energy into gas, may further strengthen interactions between the two systems and aggravate associated reliability issues \cite{guelpa2019towards,yu2018reliability,guandalini2015power,clegg2016storing}.}

\subsubsection{Integrated Optimization of Electric and Natural Gas Networks}
To ensure a power system dispatch that respects gas pipeline feasibility constraints, several studies have considered the inclusion of gas pipeline models in power grid design and operation analysis, forming methods for \emph{integrated optimization and operational planning} \cite{an03,geidl07,unsihuay07a}. %
Multiple subsequent studies on joint optimization have considered electric grid scheduling subject to security constraints that arise from the dependence on gas pipelines \cite{li08,liu09,correa14,wu11,liu10,saldarriaga13,liu11,ZhouGuWuSong2017,zlotnik17tpwrs,chen2019unit,bao2019piecewise,schwele2019coordination,badakhshan2019security}, aiming for improved power grid resilience \cite{ShaoShahidehpour2017,ZhaoConejoSioshansi2019}, better robustness against $N-1$ failures \cite{omalley2018security,HeWuLiuBie2018,jamei2018gas}, or consideration of joint expansion planning \cite{HuBieDingLin2016,ZengZhang2017,sanchez2016convex,ZhaoConejoSioshansi2018,WangQiu2017,QiuYang2016,Barati2015}.
These studies have indicated that there would be substantial economic and reliability advantages to improving coordination between the sectors \cite{jalving2017graph,pambour2018value,dokic2019security}.

\subsubsection{Information Exchange as a Barrier to Coordination}
Better coordination between natural gas and electric energy system managers can improve the reliability of both sectors. However, institutional and regulatory barriers typically prevent sharing of the necessary operational data, which is considered proprietary or competitive information. It is therefore unlikely that a single entity will be able to collect the network models and operational information required for joint optimization of both systems.
This has led to the study of coordinated operations that rely only on the limited exchange of financial and operational information \cite{zlotnik17psig,rudkevich18hicss,zhao2018shadow,zlotnik2019pipeline}, including constraints on the gas stored in pipelines \cite{clegg2015integrated, antenucci2017gas}.
Other studies have examined various coordination market mechanisms and bidding strategies \cite{yang2019coordination,chen2019equilibria,ordoudis2017exploiting, omalley2020natural}, including models that account for future energy system functions \cite{jiang2018coordinated,zexing2018coordinated}.
\avz{However, methods for coordination with limited information exchange subject to uncertainty remain limited.}

\subsection{Transient Natural Gas Pipeline Dynamics}
\avz{An important difference between the power grid and natural gas pipelines involves the time-scales of system dynamics. While power flow in transmission lines reaches steady-state within seconds,
natural gas pipelines typically never reach steady-state in intra-day operations \cite{sukharev2019phenomenological}. }
\edit{The inertia and slow dynamics of gas flow are advantageous for several reasons. Flows are constant on the second time scale of electro-mechanical transients in the power grid, leaving operators time to react to contingencies before their effects propagate through the system \cite{sukharev2017impact, omalley2018security}. 
Because of their slow dynamics and ability to store a substantial mass of gas internally, pipelines are in many ways similar in function to grid-scale batteries \cite{EPRIStorage1976}.  
The energy stored is reflected in a higher-than-minimum pressure, and is typically referred to as ``linepack''. \avz{This storage capability enables gas-fired generators to withdraw gas in excess of the scheduled rate for a limited duration without violating pipeline operating limits.}
This ability to ``pack and draft'' gas in a pipeline is captured only by dynamic models.  }

\edit{While the linepack provides short-term flexibility, the slow evolution of the gas pipeline state also implies that \avz{gas controllers must take a proactive role in predicting future system states} and take action to prevent negative impacts of potential future disturbances, utilizing metrics such as pressure cover \cite{nationalgrid2017webinar}. 
Predicting \avz{the evolution of pipeline system pressures} can be done using dynamic gas pipeline optimization or simulations, where partial differential equations (PDEs) describe how pressures, densities and mass flows evolve over time as a function of natural gas withdrawals and operation of compressor stations.
This dynamic behavior is critical to understand the impact of both generation and compressor schedules \cite{zlotnik17tpwrs,zlotnik15cdc}, renewable fluctuations \cite{chertkov15ae, chertkov15hicss}, and contingencies \cite{omalley2018security} on gas pipeline operations. 
However, because of the theoretical and computational challenges associated with PDE constrained optimization, a majority of the research on the interactions between natural gas and electric systems have utilized steady-state equations and approximations thereof, particularly when considering renewable energy uncertainty \cite{ChenWeiSunCheungSun2017,hu2019stochastic,rayati2019optimal, ordoudis2019integrated}. }

\edit{In the following, we review the steady-state and transient dynamic models of the natural gas system in more detail.}
\subsubsection{Steady-State Optimization}
Although the steady-state is typically never reached in intra-day operations, steady-state approximations, \avz{such as the} Weymouth equations \cite{borraz10phd}, are frequently utilized in the academic literature as a basis for gas pipeline modeling and optimization (including many of the above and subsequent references). Steady-state optimization of gas pipelines, utilizing the steady-state equations, has been formalized in optimal gas flow (OGF) problems \cite{wong68,misra15,riosmercado15,borraz10phd,wu17acc}. 
\avz{The constraints are non-linear algebraic equations, which are easier to solve than the time-dependent PDEs that describe transient pipeline behavior.} 
However, the steady-state models, or sequences of such models, are insufficient to capture the effect \avz{of} linepack \cite{carter03,liu11}. Furthermore, the steady-state equations (or relaxed version thereof) do not capture the phenomena needed to guarantee feasibility under dynamic conditions \cite{zlotnik17tpwrs}, and may not utilize the pipeline capacity to the greatest possible extent. 

\subsubsection{Transient Optimization}
To address the shortcomings of the steady-state natural gas models, it is important to develop optimization methods that are able to account for the transient dynamic system behavior.
Transient optimization, frequently also referred to as dynamic optimization, includes the PDE representation of the gas pipeline dynamics. These models have been expressed as model predictive optimal control problems \cite{steinbach07pde,abbaspour07,rachford09,gopalakrishnan13mpc,zlotnik15cdc,zavala14,gugat18mip,zlotnik19cdc,rudkevich19hicss,sukharev2019mathematical}, or solved using adjoint optimization \cite{omalley2018optimizing}.

The PDE constraints that represent the transient gas flows complicate the transient optimization problems \cite{moritz07phd}, and make them notoriously difficult to simulate and optimize in a tractable way \cite{rachford00,steinbach07pde,ehrhardt05}. Extension to general network structures, scalability of the computational methods, and accuracy of models and solutions continue to present challenges.  
However, recent work has made advancements towards tractable, rapid pipeline transient optimization \cite{rachford00,steinbach07pde,zlotnik15dscc,zlotnik15cdc,zavala14,gopalakrishnan13,devine14,mak16acc,mak2019dynamic}. In particular, these recent studies have resulted in modeling concepts and dynamic system representations for general large-scale pipeline systems \cite{zlotnik15dscc,dyachenko17ss}, optimal control formulations \cite{zlotnik15cdc}, algorithms based on adjoint optimization \cite{omalley2018optimizing}, comparisons of various discretization schemes \cite{mak16acc,mak2019dynamic}, extension to non-ideal gas modeling \cite{gyrya19staggered}, as well as validation of simulation and optimization approaches with respect to industrial data and commercial solvers \cite{zlotnik17psig}. \edit{Collectively, these developments can enable widespread adoption of transient gas pipeline models.}

\subsection{Uncertainty Management in Gas-Electric Energy Systems}
Both electric grids and natural gas pipeline systems experience uncertainty \avz{in} the demand and supply of energy. \avz{Flexible gas-fired generators are} frequently used to balance real-time fluctuations in the electric grid. Uncertainty and variability in  power systems thus propagates into the natural gas pipelines.
We review methods for managing this uncertainty in power grids, gas pipeline systems, and the integrated \avz{energy} system.

\subsubsection{Uncertainty in Electric Power Systems}
With increasing levels of renewable energy generation, the grid is experiencing both increased short-term variability and uncertainty. This has led to the development of uncertainty-aware optimization methods for power systems operation, in particular stochastic and robust approaches for unit commitment and optimal power flow problems.
These methods include, among many others, robust and worst-case methods \cite{panciatici2010, capitanescu2012, warrington2013, lorca2015}, two- and multi-stage stochastic programming based on samples \cite{bouffard2008, morales2009, papavasiliou2013multiarea, wang2008security}, stochastic approximation techniques \cite{tinoco2012}, and chance-constrained formulations 
\cite{ozturk2004solution, vrakopoulou2013, roald2013analytical, hamon2013value, margellos2013stochastic, bienstock2014chance, summers2014stochastic, roald2015optimal, lubin2015robust, roald2016corrective,dall2017chance, geng2019chanceconstrained}. 
Many of these methods consider the co-optimization of energy and reserve capacity.

\subsubsection{Uncertainty in Natural Gas Networks}
Increased integration of uncontrolled and variable wind generation into a power system translates into more uncertainty and variability in natural gas demand from gas-fired generating units \cite{HittingerWhitacreApt2010,chertkov15hicss,chertkov15ae}.
Variable generation schedules and reserve activation from gas-fired generators lead to a propagation of uncertainty from \avz{the power grid to gas pipelines} \cite{sahin11,sahin12}.
Several studies have investigated how this uncertainty and variability impacts gas pipelines \cite{vos2012impact,chertkov15ae,chertkov15hicss,dorsey2019effects,qiao2017interval,ChenWeiSunCheungSun2017}.  
In addition to the impact of renewable energy variability \cite{antenucci2017gas,yang2017effect}, natural gas pipelines are also influenced by uncertainty in the consumption of traditional pipeline customers \cite{tamba2018forecasting}, failures  of components such as compressors or pipelines \cite{omalley2018security,HeWuLiuBie2018,jamei2018gas}, and uncertainty regarding available \avz{pipeline} capacity \cite{ZhaoConejoSioshansi2017}.   %

\subsubsection{Integrated Uncertainty Management}
\label{ref:integrated-uncertainty}
\avz{The advantages of coordinated scheduling have inspired methods for joint management to mitigate the propagation of uncertainty from the electric grid to natural gas pipelines.}
Several studies have formulated power system scheduling problems that account for these factors. Some of these methods consider sequential optimization, where additional constraints are added to the electric scheduling problem when \avz{gas pipelines experience} constraint violations \cite{qadrdan2013operating, antenucci2017gas}, while others consider joint optimization problems. A common approach to joint optimization is to model wind generation as stochastic and uncertain, and develop two-stage stochastic formulations for joint gas and electricity system operations \cite{alabdulwahab2015stochastic, alabdulwahab2015coordination, hu2019stochastic,rayati2019optimal,ordoudis2019integrated}. Other approaches 
use interval optimization \cite{bai2016interval} or draw on robust optimization techniques \cite{liu2014look, he2016robust, bai2016robust,xiao2018two,HeWuLiuBie2018,he2018co,mirzaei2019igdt}. Most of these studies consider operational scheduling, and some include emerging power-to-gas technologies \cite{he2016robust, he2017robust, yu2018economic,zhang2018day} or expansion planning \cite{DingHuBie2018,xiao2018two,HeWuLiuBie2018}.

Solving the resulting problems can be challenging, particularly if the number of considered scenarios is large. This can be the case in scenario-based stochastic formulations, which may require a large number of scenarios to obtain accurate solutions, as well as in robust formulations that generate constraints based on, e.g., identification of worst-case scenarios \cite{he2018co}, Benders cuts \cite{alabdulwahab2015coordination}, or using column-and-constraint generation \cite{liu2014look, xiao2018two}. \edit{We also note that the method for robust co-optimization proposed in \cite{he2016robust} uses ADMM to facilitate optimization with limited information exchange.}
\edit{While the above literature provides a valuable set of methods for joint uncertainty management, and some also addresses questions related to limited information exchange, none of these methods include a transient dynamic PDE model of the natural gas flows.}
This is problematic because renewable energy uncertainty occurs at the time-scale of intra-day operation, where natural gas systems are not in steady-state. It is therefore important that the model accurately captures the impact of uncertainty on the transient gas dynamics.

\edit{In the next section, we discuss the gap identified in our review, and propose a framework to address this shortcoming.
}

\section{Framework for Coordinated Scheduling Under Uncertainty}
\label{sec:framework}
Our review highlights the benefits and challenges \avz{of} coordination between natural gas and electric systems. Successful coordination schemes \avz{require} tractable joint optimization models that also account for practical constraints such as clearly defined interfaces, as well as important physical characteristics such as time-evolution and linepack that are only captured in a transient model. Furthermore, to ensure that gas-fired generators can reliably provide balancing for intermittent renewables, it is necessary to develop a framework for coordinated scheduling under uncertainty. \edit{\emph{In our review, we \avz{note a lack of}  methods for integrated optimization and uncertainty management that also considers transient gas dynamics.}} %

\begin{figure}
    \centering
    \includegraphics[width=0.99
    \columnwidth]{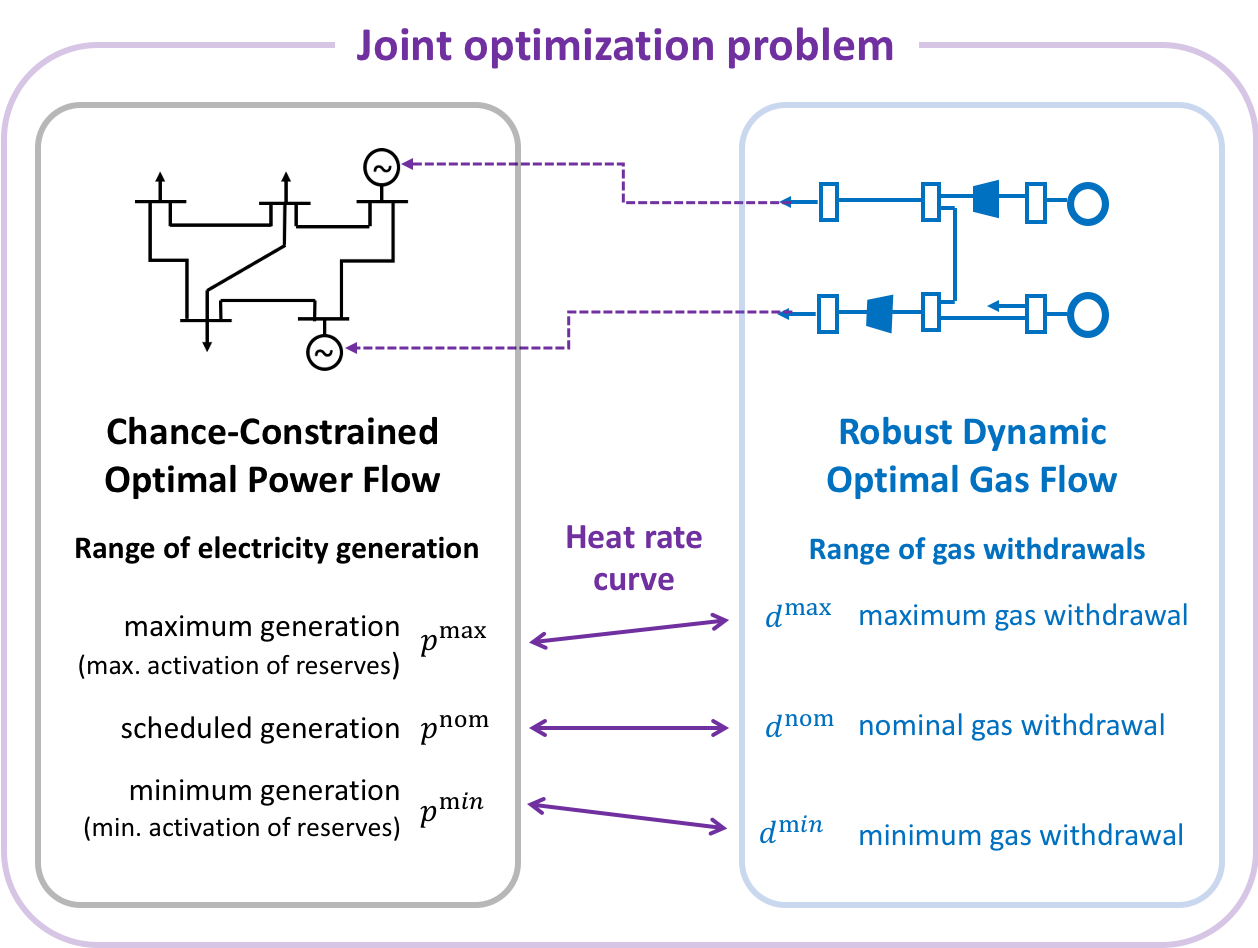}
    \caption{Schematic overview of the joint optimization framework. The framework combines chance-constrained optimal power flow with robust dynamic optimal gas flow, and determines feasible ranges of gas withdrawals from gas-fired power plants.}
    \label{fig:overview}
    \vspace{-2ex}
\end{figure}

To address this knowledge gap, we propose a practical framework for coordinated gas-electric scheduling which incorporates \edit{rigorous safety guarantees for a range of uncertain operating conditions}, a transient model of the gas pipeline, and clearly defined interfaces between the two systems. 

\edit{The proposed framework is based on the simple principle that the electric and natural gas systems agree on a range of admissible gas withdrawals for each gas generator. \avz{These ranges are formulated as maximum and minimum gas consumption profiles for each gas-fired power plant, at each point in time.
For the electric grid the maximum and minimum gas withdrawals corresponds to maximum and minimum limits on electricity generation. 
The conversion from admissible gas withdrawals to admissible electricity generation range is done using the generator heat rate curve.
Our proposed method is able to} determine maximum and minimum gas withdrawals that (i) are consistent with the needs for generation and reserve activation in the electric power grid, and (ii) guarantee robust feasibility of the gas system for any intermediate withdrawal. The three parts of the method, which are all combined in one joint optimization problem, are illustrated in Fig. \ref{fig:overview}, and summarized below.

To identify an optimal generation dispatch and reserve allocation for the electric grid, we utilize a probabilistic power system optimization method that ensures balanced power system operation and manages transmission line congestion under uncertainty.
In our framework, any multi-period stochastic or robust methods for electric grid scheduling can be utilized, as long as the scheduling algorithm provides an \emph{upper and lower bound on the gas-fired generation at each point in time}.
For our implementation in this paper, we propose to use a chance-constrained optimal power flow formulation that co-optimizes generation dispatch and reserve allocation for each generator. The chance constraints are reformulated using an analytical approach \cite{roald2013analytical, bienstock2014chance}. This method has been shown to be \avz{effective and scalable} \cite{roald2016corrective, sundar18tcst}, and relates well to existing industry practice \cite{abbaspourtorbati2016}. The details of this method are described in Section \ref{sec:power}.

The power system optimization problem is coupled with the gas system optimization problem through the \emph{heat rate curve}, which transforms 
the nominal, maximum and minimum generation outputs (corresponding to scheduled generation and full activation of upwards or downwards reserves, respectively) into nominal, maximum and minimum gas withdrawals for each generator.  These \emph{uncertain, but bounded gas withdrawals} are combined into three scenarios for gas \avz{pipeline} operation corresponding to 1) all generators operating at nominal generation, 2) all generators \avz{consuming} the maximum amount of gas, and 3) all generators \avz{consuming} the minimum amount of gas. Together, these three scenarios describe the interface between the two systems. Details of how the integration of the natural gas and electric models are given in Section \ref{sec:integration}.

To guarantee robust feasibility for the gas system, we utilize recent results for gas flow monotonicity \cite{vuffray15cdc,misra2016monotone,misra20pieee} to formulate a \emph{robust transient gas flow optimization}. This step is an important novelty of our proposed method, as we are able to guarantee feasibility for a wide range of practical operating scenarios, while considering no more than three scenarios within the optimization problem. The formulation of the transient dynamic gas optimization and the extension to a robust formulation are described in Section \ref{sec:gas}.}

\avz{The main advantages of the proposed method are:}
\begin{itemize}
    \item \avz{By guaranteeing that any gas withdrawal within the determined maximum and minimum bounds
    is feasible, we allow gas-fired generators to vary their electricity production to balance loads in the power system, without risking curtailments in the gas supply or cause constraint violations in the gas pipeline system.}
    \item \avz{Feasibility is guaranteed for a wide range of operating scenarios in the coupled gas-electric system, while only considering three scenarios for the natural gas withdrawals.} 
    \item \avz{The a small number of required scenarios enables tractable optimization problems, even when considering the transient dynamic gas flows.}
\end{itemize}
\avz{Our method expands upon existing literature by considering both transient dynamics and uncertainty in an integrated optimization problem. Based on our review, addressing these aspects together is nontrivial and novel. To our knowledge, previous studies on joint optimization under uncertainty \cite{alabdulwahab2015stochastic, alabdulwahab2015coordination, bai2016interval, bai2016robust,
DingHuBie2018, HeWuLiuBie2018, he2018co, he2016robust, he2017robust, hu2019stochastic, liu2014look, mirzaei2019igdt,
ordoudis2019integrated, rayati2019optimal, xiao2018two, yu2018economic, zhang2018day} consider only steady-state gas models.} %

The details of the mathematical model, which considers the chance-constrained optimal power flow and robust dynamic gas flow problem as one integrated optimization problem, are given in the following sections. 
\edit{In the interest of clarity, we first separately introduce the models for the electric power system and the natural gas network, before combining them into an integrated model. We first}
start with the chance-constrained power system model, \edit{where we discuss reserve scheduling in detail. We then} move on to \edit{gas pipeline analysis, where we first describe the deterministic transient optimization model, and then discuss our extensions to a robust optimization formulation.} Finally, \edit{we discuss how we} model the interactions \edit{between the systems using the heat rate curve, and formulate the proposed integrated optimization for gas-electric systems.}

\section{Chance-Constrained Optimal Power Flow for Generation and Reserve Scheduling}
\label{sec:power}
In this section we describe a method for probabilistic optimization of electric transmission systems with high penetrations of renewable energy, which guarantees constraint satisfaction with a high probability. This guarantee is naturally enforced using chance constraints, giving rise to a multi-period, chance-constrained optimal power flow problem. While the formulation is mostly a review of pre-existing work  \cite{roald2013analytical, bienstock2014chance, roald2016corrective, sundar18tcst}, we include a more in-depth discussion of reserve scheduling. Using chance constraints for reserve scheduling aligns well with probabilistic \avz{criteria} for reserve dimensioning used within ENTSO-E \cite{ENTSOE2012, abbaspourtorbati2016}, which requires the reserve capacities to be sufficiently large to cover imbalances with a high probability. 
For example, the Swiss market for ancillary services \cite{abbaspourtorbati2016} constructs an empirical probability distribution based on historical power imbalances and enforces a chance constraint in the market clearing to ensure that the procured reserve capacities will exceed the required reserve activation with a predetermined probability. Our approach formalizes this method for reserve scheduling, and includes probabilistic feasibility guarantees for line flows as well.

\subsection{Power System Network Modeling}
Let $G_P = (\mathcal{V}_P, \mathcal{E}_P)$ denote the graph of the power transmission network, where $\mathcal{V}_P$ is the set of nodes (i.e., buses) with $|\mathcal{V}_P| = m$ and $\mathcal{E}_P$ is the set of edges (i.e., transmission lines) of the system with $|\mathcal{E}_P| = n$. All quantities associated with a line from bus $i$ to bus $j$ are identified by subscript $ij$. 
Let $\mathcal{G}$ denote the set of generators in the network. For \avz{simplicity} of notation, we assume that there is exactly one generator with production $p_i(t)$ and one load with consumption $h_i(t)$ per node, such that $|\mathcal{G}|=|\mathcal{V}_P| = m$. 

\subsubsection{Modeling Uncertain Demand}
The source of uncertainty in our model is power injection uncertainty, i.e., how much power will be injected into or withdrawn from the system by uncontrollable sources such as renewable energy generators or loads. 

We denote the forecasted demand for power at each node $h_i(t)$ as continuous demand functions defined for $0\leq t\leq T$ where $T=24$ hours.
To model uncertainty regarding the actual demand, we express each power injection in the system as
\begin{align}
    \tilde{h}(t) = h(t) - \omega(t).  \label{eq:load_policy}
\end{align}
Here, $\tilde{h}(t)$ denotes the vector of all real-time loads in the system at time $t$. The vector $h(t)$ is the forecasted system load, and $\omega(t)$ correspond to the \edit{difference between the forecasted and actual load. Since the actual load is not known until real-time operations, the exact value of $\omega(t)$ is unknown and we will refer to $\omega(t)$} as the uncertain part \avz{nodal demand}. Note that $\omega(t)$ can represent uncertainty arising from either load or renewable energy fluctuations, or a combination of both. \avz{We also assume that the covariance matrix $\Sigma(t)$ of $\omega(t)$ is known and finite.}
\edit{The total difference between forecasted and actual load is referred to as the total real-time power imbalance, which we denote by $\Omega(t)$ and define as}
\begin{align*}
    \Omega(t) = \sum_{i\in\mathcal{V}_P} \omega_i(t).
\end{align*}

\subsubsection{Power Balancing in Real-Time Operations}
\edit{An important aspect of power system operations is to maintain balance between the power produced by the generators and the power consumed by the loads at all times. To ensure that power balance is maintained in real-time operation, the changes in the total load, represented by the imbalance $\Omega(t)$, are compensated through the activation of reserves. In this process, conventional generators are asked to either increase or decrease their power generation to match the power consumption of loads in real time.} Following standard conventions for power system balancing implemented in, e.g., the automatic generation control (AGC), we assume that activation of reserves by individual generators is given by a vector of participation factors $\beta$ that describe how much each generator contributes to compensate for the imbalance $\Omega(t)$. The participation factors give rise to a policy for generation control of the form
\begin{equation}
    \tilde{p}(t) = \generationsetpoint(t) - \beta(t)\Omega(t)  \label{eq:gen_policy}.
\end{equation}
In this policy, $\generationsetpoint(t)$ represents the vector of scheduled power generation, while the term $-\beta(t) \Omega(t)$ represents real-time adjustments in the output, referred to as \emph{activation of reserves}. 

Assuming a lossless DC power flow approximation of the system, total power balance requires that the total generation equals total demand for any realization $\omega(t)$, i.e.,
\begin{align}
    &\textstyle \sum_{i\in\mathcal{G}}\left(p_i(t) - \beta_i\Omega(t)\right) - \sum_{j\in\mathcal{V}_P}\big(h_j(t) - \omega_j(t)\big) = 0.
\end{align}
This constraint can be rearranged into the expression
\begin{align}
    &{\textstyle\sum_{i\in\mathcal{G}}p_i(t) - \sum_{j\in\mathcal{V}_P}h_j(t) + 
    \left(1-\sum_{i\in\mathcal{G}}\beta_i\right) \Omega(t) = 0.} \nonumber
\end{align}
Here, the first two terms enforce power balance for the forecasted operating point, while the last term ensures power balance during fluctuations. 
To guarantee power balance for any realization of $\Omega(t)$, we enforce the conditions
\begin{subequations} \label{eq:balancing}
\begin{align}
    &\sum_{i\in\mathcal{G}}p_i(t) - \sum_{j\in\mathcal{V}_P}h_j(t) = 0, \quad &&\forall ~t\in T, \\
    &\sum_{i\in\mathcal{G}}\beta_i(t) = 1, \quad \beta_i(t)\geq 0\quad&&\forall~i\in\mathcal{G},~t\in T.
\end{align}
\end{subequations}
In some power systems, the participation factors $\beta$ remain fixed for prolonged periods of time. We will assume that they can be updated at every scheduling interval (e.g., for every hour in the day-ahead market clearing), and that they are co-optimized along with the procurement of reserves.

\subsection{Reserve Scheduling as a Chance Constraint}
Reserve scheduling determines the reserve capacities $r^+_i$ and $r^-_i$ for each generator $i$. Probabilistic reserve scheduling ensures that \edit{(i)} the generators have the physical capacity required to provide reserve activation, and \edit{(ii)} the total up-reserve and down-reserve capacities $r^+$ and $r^-$ are sufficiently large to make up for the power imbalances. Since the real-time power unbalance $\Omega(t)$ is a random variable, these requirements cannot be enforced as deterministic constraints. Instead, we follow industry practice \cite{abbaspourtorbati2016} and previous results \cite{roald2016corrective} to ensure that the reserve capacity is sufficient with a \emph{high probability}.

\subsubsection{Reserve Sufficiency for Individual Generators}
To ensure that each individual generator $i\in\mathcal{G}$ has sufficient capacity available to provide reserve activation, we formulate the generator constraints as
\begin{align}
&\begin{matrix}
\generationsetpoint_i(t) + r_i ^+(t) \leq p_i^{\max}, ~~\forall~i\in\mathcal{G},~t \in \mathcal T \\
\generationsetpoint_i(t) - r_i ^-(t) \geq p_i ^{\min}, ~~\forall~i\in\mathcal{G},~t \in \mathcal T %
\end{matrix} \,\,  \label{eq:gen_cap} \\[2pt]
&\begin{matrix}
\mathbb{P}_\Omega(-\beta_i(t)\Omega(t) \leq ~~r_i ^+(t)) \geq 1-\varepsilon,  ~~\forall~i\in\mathcal{G},~t \in \mathcal T\\ %
\mathbb{P}_\Omega(-\beta_i(t)\Omega(t) \geq -r_i ^-(t)) \geq 1-\varepsilon, ~~ \forall~i\in\mathcal{G},~t \in \mathcal T%
\end{matrix} \,\,  \label{eq:gen_res} \\
&r_i(t)^+ \geq 0, ~~ r_i(t)^- \geq 0,  ~~ \forall~i\in\mathcal{G},~t \in \mathcal T \label{eq:pos_res}
\end{align}
The first two constraints \eqref{eq:gen_cap} ensure that the combination of the scheduled generation $p_i(t)$ and reserve capacities $r_i^+, r_i^-$ remain within the upper and lower generation limits $p_i^{\max}, p_i^{\min}$. 
The latter two constraints \eqref{eq:gen_res} enforce that the activated reserves do not exceed the scheduled reserve capacities $r_i^+, r_i^-$ with probability $1-\varepsilon$, for each generator $i$.
These type of constraints are referred to as \emph{individual chance constraints}, as they consider each generator individually.
Eq. \eqref{eq:pos_res} ensures that the reserve capacities are positive.

Note that the reserve constraints in \eqref{eq:gen_res} are a chance constrained version of \emph{hard} constraints on generation output, which is \emph{directly controllable}. 
In reality, a generator may not alter its generation by more than the procured reserve capacity (and certainly never produce beyond the maximum generation capacity $p^{\max}$, which is physically impossible). A chance-constraint violation therefore does not imply a generator overload, but rather describes the probability that the simple affine policy for reserve activation described in \ref{eq:gen_policy} is no longer working. Instead, the system operator must resort to alternative controls to maintain power balance. This setting has been \avz{studied in detail} \cite{kannan2019stochastic}. %

In addition to the above constraints, the problem must take into account of fast the generators are able to modulate their output. To ensure that the generators scheduled for reserves are physically capable of delivering them, system operators typically test generators to ensure that they are able to follow reserve activation signals. The reserves procured from each generator must lie within the maximum capacity that the generator has been certified for: 
\begin{subequations}    \label{eq:reserve_bid_bounds}
\begin{align}
    & 0 \leq r^+_i(t) \leq r^{+}_{i,\max}, \label{maxUpRes}\\
    & 0 \leq r^-_i(t) \leq r^{-}_{i,\max},\quad \forall i \in \mathcal{G}, t \in \mathcal T, \label{maxDownRes}
\end{align}
\end{subequations}
Here, $r^{+}_{i,\max}, r^{-}_{i,\max}$ denote the maximum up and down reserve capacity of generator $i$. Finally, the change in generation set points between two consecutive time periods must lie within the ramping limits of the generator,
\begin{subequations}    \label{eq:ramping_limits}
\begin{align}
    & \generationsetpoint_i(t)-\generationsetpoint_i(t+1)\leq R^{\max}_i, \\
    & \generationsetpoint_i(t+1)-\generationsetpoint_i(t)\leq R^{\max}_i, \ \forall i \in \mathcal{G}, t \in \mathcal T. 
\end{align}
\end{subequations}

\subsubsection{Reserve Sufficiency for the Overall System}
While each generator needs sufficient capacity to provide the requested reserves, it is also more important to ensure that the total amount of scheduled reserve capacity from all generators is sufficient for balancing. This can be expressed as the following \emph{joint chance constraint},
\begin{align}
&\mathbb{P}_\Omega
\begin{pmatrix} 
-\beta_i(t)\Omega(t) \leq ~r_i ^+(t), \, \forall~i\in\mathcal{G} \\
-\beta_i(t)\Omega(t) \geq -r_i ^-(t), \, \forall~i\in\mathcal{G}
\end{pmatrix} 
\geq 1-\varepsilon_{joint}, \, \forall \, t \in \mathcal T \label{eq:res_joint} 
\end{align}
This constraint guarantees that the reserve activation is within the scheduled reserve capacities for \emph{all} generators, with probability $1-\varepsilon_{joint}$. This is similar to requirements imposed by system operators, such as Swissgrid \cite{abbaspourtorbati2016}.

In general, solving problems with joint chance-constraints like \eqref{eq:res_joint} is more challenging than solving problems with individual chance constraints like \eqref{eq:gen_res}. However, using the particular structure of \eqref{eq:res_joint}, it has been shown \cite{roald2016corrective} that enforcing the single chance constraints \eqref{eq:gen_res} gives strong guarantees on the satisfaction of the joint reserve requirement \eqref{eq:res_joint}. We derive these results here in more detail.

Because the total power imbalance $\Omega$ is a scalar random variable,
the individual chance constraints \eqref{eq:gen_res} impose a quantile constraint for each generator $i \in \mathcal{{G}}$:
\begin{align}
    &\mathbb{P}_\Omega\Big(\!-\!\beta_i(t)\Omega(t) \leq r_i ^+(t)\Big) \!\geq\! 1-\varepsilon \ \equiv \  -\beta_i(t)Q_{\Omega}(\varepsilon)\leq r_i ^+(t), \nonumber\\
    &\mathbb{P}_\Omega\Big(\beta_i(t)\Omega(t) \leq r_i ^-(t)\Big) \!\geq\! 1-\varepsilon \ \equiv \  \beta_i(t)Q_{\Omega}(1-\varepsilon)\leq r_i^{-}(t) \nonumber
\end{align}
where $Q_\Omega(\varepsilon)$ denotes the $\varepsilon$-quantile of the random variable $\Omega$. Note that while we have two constraints for each generator, the values of $Q_\Omega(\varepsilon)$ and $Q_\Omega(1-\varepsilon)$ are shared among all the constraints. Therefore, the above constraints are equivalent to \edit{requiring that the ratio between the reserve capacities $r^+_i, r^-_i$ and the participation factors $\beta_i$ for all generators satisfy  }
\begin{align}    \label{eq:joint_equivalent}
    \max_{i \in \mathcal{G}} -\frac{r_i ^+(t)}{\beta_i(t)} \leq Q_{\Omega}(\varepsilon) \leq Q_{\Omega}(1-\varepsilon) \leq  \min_{i \in \mathcal{G}}\frac{r_i ^-(t)}{\beta_i(t)}.
\end{align}
As a result, \eqref{eq:gen_res} are satisfied for all $i \in \mathcal{G}$ if and only if \eqref{eq:joint_equivalent} is satisfied. Further, by definition of the quantile, \eqref{eq:joint_equivalent} is satisfied if and only if
\begin{align}   %
   &\pW\Big(\!-\!r_i ^+(t) \leq \beta_i(t)\Omega(t) \leq r_i ^-(t), ~\forall i\!\in\mathcal{G}, t \in \mathcal\! T\Big) \!\geq\! 1\!-\!2\varepsilon.\nonumber
\end{align}
which is equivalent to the joint chance constraint \eqref{eq:res_joint} with $\varepsilon_{joint}=2\varepsilon$.
Therefore, the individual local chance constraints we impose in \eqref{eq:gen_res} automatically enforce the global joint chance constraint in \eqref{eq:res_joint}.

\subsection{Transmission Line Constraints}
\label{sec:integrated}
With the DC power flow model, the power flows can be expressed as a linear function of the nodal power injections using the power transfer distribution factors (PTDF) \cite{christie2000},
\begin{align} \label{eq:ptdf}
    p^{line}_{\edit{mn}}(t) = M_{(\edit{mn},\cdot)}\left(\generationsetpoint(t)-\beta(t)\Omega(t)-h(t)+{\omega}(t)\right),
\end{align}
\edit{where $p^{line}_{mn}$ denotes the line flow on the transmission line between nodes $m$ and $n$, $M$ is the PTDF matrix as defined in \cite{vrakopoulou2013} and $M_{(mn,\cdot)}$ denotes the row of $M$ corresponding to line $mn$.} 
This model captures the impact of the uncertain fluctuations and reserve activation on the power flows $p^{line}_{mn}(t),~\forall {mn}\in\mathcal{E}_P$, which themselves become random variables.
Similar to the generation constraints, we take a probabilistic approach and enforce the power flow limits using chance constraints, %
\edit{
\begin{subequations} \label{eq:cc_lineflow}
\begin{align}
  & \pw\left(p^{line}_{mn}(t) \leq ~~p_{mn}^{\max}\right) \geq  1-\epsilon_{mn}, \quad\forall~mn \in \mathcal{L} \label{upperlinechance}\\
  & \pw\left(p^{line}_{mn}(t) \geq -p_{mn}^{\max}\right) \geq  1-\epsilon_{mn},  \quad\forall~mn \in \mathcal{L}
\label{lowerlinechance}
\end{align}
\end{subequations}}
\noindent with $p^{line}_{\edit{mn}}$ defined as in \eqref{eq:ptdf}. \edit{Here, \eqref{upperlinechance} and \eqref{lowerlinechance} enforce the upper and lower bounds on the line flow (corresponding to the maximum line flow in both directions).}

The probability of violation in \eqref{eq:cc_lineflow} has a different meaning than for the generator constraints in \eqref{eq:gen_res}.
Whereas the generator constraints \eqref{eq:gen_res} are \emph{hard} constraints on variables that are directly controllable, the chance constraints on the power flow \eqref{eq:cc_lineflow} correspond to \emph{soft} constraints on the flow on each transmission line, which is only \emph{indirectly controllable}. For these chance constraints, a constraint violation implies that a line might experience an actual, physical overload with probability $1-\varepsilon$. Depending on the size and duration of the overload, the operator can either wait for the overload to resolve itself (if it is small and short-lived) or implement alternative control actions, such as manual redispatch, to relieve line congestion. 

For the generators, the individual chance constraints (imposed separately on each generator) imply joint satisfaction of the chance constraint. For the line flow constraints \eqref{eq:cc_lineflow}, this result does not hold, as the constraints depend not only on the total power imbalance $\Omega(t)$, but on the full random vector $\omega(t)$. Therefore, the probability that any line is overloaded is, in the worst case, equal to the number of lines times their individual violation probability, i.e. $\varepsilon_{max}=\max\{n\cdot\epsilon,1\}$. However, given the structure of power grids, the number of line constraints that experiences violations is typically very low \cite{bienstock2014chance,roald2016corrective}. It is therefore possible to very effectively control the joint violation probability using individual chance constraints on the lines.

\subsection{Reformulation of the Chance Constraints}
The chance-constrained optimization problem \eqref{eq:cc_opf} can only be solved after reformulation as a deterministic optimization problem. While there are multiple approaches for the chance constraint reformulation, we utilize the analytical reformulation from \cite{roald2013analytical,bienstock2014chance} which is highly scalable \cite{bienstock2014chance,roald2016corrective}. For simplicity, we assume that the forecast errors follows a multivariate Gaussian distribution. However, the same formulation can be extended to account for more general or partially unknown distribution \cite{roaldArxiv} as well as general elliptical uncertainty sets \cite{ben2009robust}. 

\edit{Specifically, we assume that the uncertainty $\omega(t)$ follow a multivariate Gaussian distribution with a known covariance matrix $\Sigma$ for each time step. For simplicity of notation, we ignore the time-dependency of $\Sigma$ and assume that $\Sigma$ is constant throughout the time horizon. However, in reality the constraints could easily be changed to account for a separate $\Sigma$ at each time step. We assume that we have unbiased forecasts of the load, such that the expected value of $\omega(t)$ is $0$.}
With the analytical reformulation, the chance constraints \eqref{eq:gen_res} \edit{for each generator $i$} can be reformulated as linear constraints \cite{li2015analytical},
\begin{subequations}\label{eq:gen_res_ref}
\begin{align}
&r_{\edit{i}}^+ + \beta_{\edit{i}}Q_{\Omega}(1-\epsilon) \geq 0, \quad \edit{\forall i \in \mathcal{G}}\label{eq:gen_res_up}\\
&r_{\edit{i}}^- - \beta_{\edit{i}}Q_{\Omega}(1-\epsilon) \geq 0,  \quad \edit{\forall i \in \mathcal{G}} \label{eq:gen_res_down}
\end{align}
\end{subequations}
\edit{where the quantile $Q_{\Omega}(1-\epsilon)= \Phi^{-1}(1-\epsilon)\sqrt{\boldsymbol{1}^T\Sigma\boldsymbol{1}}$ is a constant. Here, $\Phi^{-1}(1-\epsilon)$ represents the inverse cumulative distribution function of the standard normal distribution evaluated at $1-\varepsilon$ (i.e., the $1-\epsilon$ quantile) and $\boldsymbol{1}$ is a vector with entries of one.} 
The transmission line constraints \eqref{eq:cc_lineflow} for each line \edit{$mn$ are reformulated} as second-order cone constraints \cite{bienstock2014chance},
\begin{subequations}\label{eq:line_ref}
\begin{align}
& M_{(\edit{mn},\cdot)}\left(\generationsetpoint(t)-h(t)\right) + \Phi^{-1}(1-\epsilon)~\!s_{\edit{mn}} \leq p_{\edit{mn}}^{\max}, \label{eq:upper_line}\\
& M_{(\edit{mn},\cdot)}\left(\generationsetpoint(t)-h(t)\right) - \Phi^{-1}(1-\epsilon)~\!s_{\edit{mn}} \geq -p_{\edit{mn}}^{\max} ,\label{eq:lower_line}\\
&\text{with }{\textstyle s_{\edit{mn}} = \sqrt{(M_{(\edit{mn},\cdot)}(I-\beta \mathbf{1}))\Sigma (M_{(\edit{mn},\cdot)}(I-\beta \mathbf{1}))^T}}. \nonumber
\end{align}
\end{subequations}
Here, $s_{\edit{mn}}$ is a variable representing the standard deviation of the line flows,\edit{ $I$ is the identity matrix and $\boldsymbol{1}$ is the vector with entries of one. For more details on the derivation of these equations, we refer the reader to \cite{roald2016corrective}.}%

\subsection{Multi-Period Chance-Constrained Optimal Power Flow}  %

Based on the modeling considerations presented in the previous section, we formalize the generation and reserve scheduling problem as a multi-period, chance-constrained optimal power flow (OPF) problem. 

\subsubsection{Decision variables}
Our chance-constrained OPF model includes the following decision variables,
\begin{subequations}    \label{eq:decision_variables_power}
\begin{align}
    &\mbox{Generation set points:} \quad &&\mathbf{\generationsetpoint} \equiv \big(\generationsetpoint(t), \ t \in \mathcal T\big),   \nonumber\\
    &\mbox{Up and down reserves:} \quad &&\mathbf{r} \equiv \big(r^+(t),r^-(t) \ t \in \mathcal T\big),  \nonumber\\
    &\mbox{Participation factors:} \quad &&{\boldsymbol\beta} \equiv \big(\beta(t) \ t \in \mathcal T\big). \nonumber
\end{align}
\end{subequations}

\subsubsection{Probabilistic Power Flow Constraints}
For ease of exposition, we define the \emph{constraint set} $\Pi$ that a given choice of decision variables $\mathbf{\generationsetpoint(t),r,\beta(t)}$ must satisfy:
\begin{subequations}    %
\begin{align}
&\Pi(\mathbf{\generationsetpoint},\mathbf{r},\boldsymbol{\beta}) = \nonumber\\
&\quad \begin{cases}
& \mbox{Total power balance}~\eqref{eq:balancing}, \nonumber \\
& \mbox{Generator limits}~\eqref{eq:gen_cap}, \nonumber \\
& \mbox{Probabilistic reserve requirement}~\eqref{eq:gen_res_ref}, \nonumber \\
& \mbox{Limits on reserve procurement}~\eqref{eq:reserve_bid_bounds},\\
& \mbox{Generator ramping limits}~\eqref{eq:ramping_limits}, \\
& \mbox{Line flow constraints}~\eqref{eq:ptdf},\eqref{eq:line_ref} 
\end{cases}
\end{align}
\end{subequations}

\subsubsection{Objective function}

The objective function for the power system evaluates the cost of generation and reserves:
\begin{align}
    J_P(\mathbf{\generationsetpoint},\mathbf{r}) &= \sum_{i\in \mathcal G} \sum_{i\in \mathcal T} \bigl\{K_ip_i(t) \bigr\} \nonumber \\
    & \,\, + \sum_{i\in \mathcal G} \sum_{i\in \mathcal T} \bigl\{a_i (r_i^+(t)+r_i^-(t))\bigr\} \label{eq:cost_power}
\end{align}
Here, $K_i$ and $a_i$ represent the cost of generation and reserve capacity, respectively.

\subsubsection{Multi-period chance-constrained DC OPF} 
The MP-CCOPF can be compactly formulated as follows:
\begin{subequations} \label{eq:cc_opf}
\begin{align}
    &\!\!\min\limits_{\mathbf{\generationsetpoint},\mathbf{r},\boldsymbol{\beta}}&& \mbox{Generation and Reserve Cost: $J_P(\mathbf{\generationsetpoint},\mathbf{r})$} \\
&\mbox{s.t.: \,\,\,}
&& \mbox{Power flow constraints:} ~\Pi(\mathbf{\generationsetpoint},\mathbf{r},\boldsymbol{\beta}).
\end{align}
\end{subequations}

\section{Robust Optimization for \\ Transient Gas Pipeline Scheduling} \label{sec:gas}

In this section we provide an overview of a standardized approach for modeling large-scale gas transmission pipelines for optimal control, which has been applied in several studies \cite{zlotnik15cdc,zlotnik17tpwrs,sundar18tcst,carreno2019adversarial}, and also validated with respect to a hydraulic model and sensor data of a pipeline system \cite{zlotnik17psig}. We also review several monotone system theorems that apply to the dynamic equations we examine here, and then utilize these results to formulate \edit{a novel} robust gas pipeline flow scheduling problem, \edit{which is essential for} the subsequent joint optimization formulation.

\subsection{Modeling Gas Transmission Pipeline Dynamics}

A large-scale gas transmission network can be modeled for transient analysis as a set of edges (which represent pipes) that are connected at nodes (which represent junctions).  The topology of the network is described as a connected directed metric graph $(\mathcal V, \mathcal E)$ where $\mathcal V$ and $\mathcal E$ denote the sets of nodes and edges, respectively, where $(i,j) \in \mathcal E$ represents an edge that connects nodes $i, j\in \mathcal V$. The dynamics of the entire system are given by characterizing the gas flow physics on each edge, as well as compatibility conditions for each node.

\subsubsection{Gas Flow Modeling} \edit{There is a well-established consensus that compressible gas flow in a pipe in the ideal gas regime is well-described using the one-dimensional isothermal Euler equations \cite{wylie78,osiadacz87,thorley87}:
\begin{subequations} \label{eq:gaspde0}
\begin{align}
    \frac{\partial \rho}{\partial t} + \frac{\partial (\rho u)}{\partial x} & = 0 \label{eq:gaspde0a} \\
    \frac{\partial (\rho u)}{\partial t} + \frac{\partial (\pi + \rho u^2)}{\partial x} & = - \frac{\lambda}{2D}\rho u |u| - \rho g \frac{\partial h}{\partial x} \label{eq:gaspde0b} \\
    \pi = \rho ZR\mathbf{T} & = a^2 \rho \label{eq:gaspde0c}
\end{align}
\end{subequations}
Equations \eqref{eq:gaspde0} represent mass conservation, momentum conservation, and the gas equation of state law.  The state variables $u$, $\pi$, and $\rho$ represent gas velocity, pressure, and density, respectively, and depend on time $t\in[0,T]$ and space $x\in(0,L)$, where $T$ is a finite time horizon and where $L$ is the length of the pipe, and the variable $h$ gives the elevation of the pipeline.  The dimensionless parameter $\lambda$ is the friction factor that scales the phenomenological Darcy-Weisbach term that models momentum loss caused by turbulent friction.  Other parameters are the internal pipe diameter $D$, and the wave (sound) speed  $a=\sqrt{ZR\mathbf{T}}$ in the gas where $Z$, $R$, and $\mathbf{T}$ are the gas compressibility factor, specific gas constant, and absolute temperature, respectively, and the gravitational acceleration constant $g$.   In our study, we assume that gas pressure $\pi$ and gas density $\rho$ satisfy the ideal gas equation of state where the wave speed in \eqref{eq:gaspde0c} is constant. Although in practice non-ideal modeling is necessary to correctly represent flows at pressures used in mainline transmission pipelines, ideal gas modeling does not qualitatively change the obtained results, so we adopt it for the sake of simplicity in the exposition of the problem formulation.  The extension to non-ideal gas modeling can be obtained through the application of a nonlinear transform \cite{gyrya19staggered}.   

The terms $\partial(\rho u^2)/\partial x$ and $\partial(\rho u)/\partial t$ in equation \eqref{eq:gaspde0b} represent kinetic energy and inertia, respectively.  It is standard to apply the transformation of variables by defining the per area mass flux $\varphi=\rho u$.  Baseline assumptions for gas transmission pipelines are \cite{herty10}: gas flow is an isothermal process; all pipes are horizontal and have uniform diameter and internal surface roughness; flow is turbulent and has high Reynolds number; and the flow process is adiabatic, i.e. there is no heat exchange with ground. With these assumptions, the coefficients $R$, $\mathbf{T}$, $D$, and $\lambda$ can be approximated by constants, and the equations \eqref{eq:gaspde0} can be reduced to}
\edit{
\begin{subequations} \label{eq:gaspde1}
\begin{align} 
    \frac{\partial \rho}{\partial t} + \frac{\partial \phi}{\partial x} & = 0 \\
    \frac{\partial \phi}{\partial t} + \frac{\partial \pi}{\partial x} & = - \frac{\lambda}{2D}\frac{\phi|\phi|}{\rho} \label{eq:gaspde1:b} \\
    \pi & = a^2 \rho.  \label{eq:gaspde1:c} 
\end{align}
\end{subequations}}
\edit{There exists a consensus in the literature regarding these initial assumptions. There is controversy about the inertial term $\partial \phi/\partial t$ in \ref{eq:gaspde1:b}, which is often omitted for mathematical convenience in optimization of transient pipeline flows.  In general, the term can be neglected when the transients are slow, as shown in empirical studies \cite{osiadacz84,gyrya19staggered}, and which we verify by simulations in our concurrent study \cite{misra20pieee}.  There is substantial precedent for applying this approximation to represent gas flow dynamics in pipeline systems in the physical regime of normal operations \cite{chaudry08,osiadacz87}.

On each pipe $(i,j)\in\mathcal E$, we therefore describe the gas flow and density dynamics using a further simplification of the equations \eqref{eq:gaspde1} assuming the ideal gas law and dropping the inertial flux derivative term.  There is a well-established consensus that this representation is sufficient to qualitatively represent dynamics on the time-scales relevant for modeling the regime of normal pipeline operation \cite{osiadacz84,herty10}, assuming a horizontal pipe with gas at constant temperature and slow boundary transients that do not cause waves or shocks.
The simplified equations are given by }
\begin{subequations}
\label{eq:pde_1}
\begin{align}
& \partial_t \rho_{ij} + \partial_x \varphi_{ij} = 0, \label{eq:pde_1a}\\
& a^2 \partial_x \rho_{ij} = -\frac{\lambda_{ij}}{2D_{ij}} \frac{\varphi_{ij} |\varphi_{ij}|}{\rho_{ij}}. & \label{eq:pde_1b}
\end{align}
\end{subequations}
Here, the variables $\rho_{ij}$ and $\varphi_{ij}$ denote the instantaneous gas density and mass flux (in per-area units), and are defined on the domain $[0,L_{ij}] \times \mathcal [0,T]$ where $L_{ij}$ is the pipe length. 
The nonlinear term on the right hand side of \eqref{eq:pde_1b} aggregates friction effects, and the parameters there are the Darcy-Wiesbach friction factor $\lambda_{ij}$ and pipe diameter $D_{ij}$.  
We henceforth will denote the pipe cross-sectional area by $X_{ij}$.  
The sign of $\varphi_{ij}$ indicates flow direction, and we may write $\varphi_{ij}(x_{ij},t)=-\varphi_{ji}(L_{ij}-x_{ij},t)$ to express the flow in the reverse direction.   
We define densities and flows at edge domain boundaries by
\begin{subequations}
\begin{flalign}
& \underline{\rho}_{ij}(t) \triangleq \rho_{ij}(t, 0), \quad \bar{\rho}_{ij}(t) \triangleq \rho_{ij}(t, L_{ij}), & \nonumber\\
& \underline{\varphi}_{ij}(t) \triangleq \varphi_{ij}(t, 0), \quad \bar{\varphi}_{ij}(t) \triangleq \varphi_{ij}(t, L_{ij}), & \nonumber%
\end{flalign}
\end{subequations}
The equation \eqref{eq:pde_1} has a unique solution given initial conditions and one boundary condition at each end of the pipe.  

\subsubsection{Nodal Compatibility Conditions} 
Each node $i \in \mathcal V$ is associated with a time-dependent nodal gas density $\rho_i(t) \geq 0$ and a gas withdrawal with a mass flow rate $d_j(t)$. Here, $d_j(t)\geq0$ and $s_j(t)\leq0$ denote consumption and supply, respectively.
All nodes $j\in\mathcal V$ are subject to physical flow balance, given by
\begin{equation}
    d_j(t) + s_j(t)=\sum_{i\in\mathcal \partial_+j}X_{ij} \overline{\varphi}_{ij}(t)- \sum_{k\in\mathcal \partial_-j}X_{jk}\underline{\varphi}_{jk}(t), \,\forall\, j\in\mathcal V_d. \label{eq:massflowbal1}
\end{equation}
The notations $\partial_{+}j=\left\{ i\in \cV\mid(i,j)\in \cE\right\}$ and $\partial_{-}j=\left\{ k\in \cV\mid(j,k)\in \cE\right\} \subset \cV$ are used  to distinguish between flows into and out from the node.
\begin{figure}[t]
\centering
\includegraphics[width=.95\linewidth]{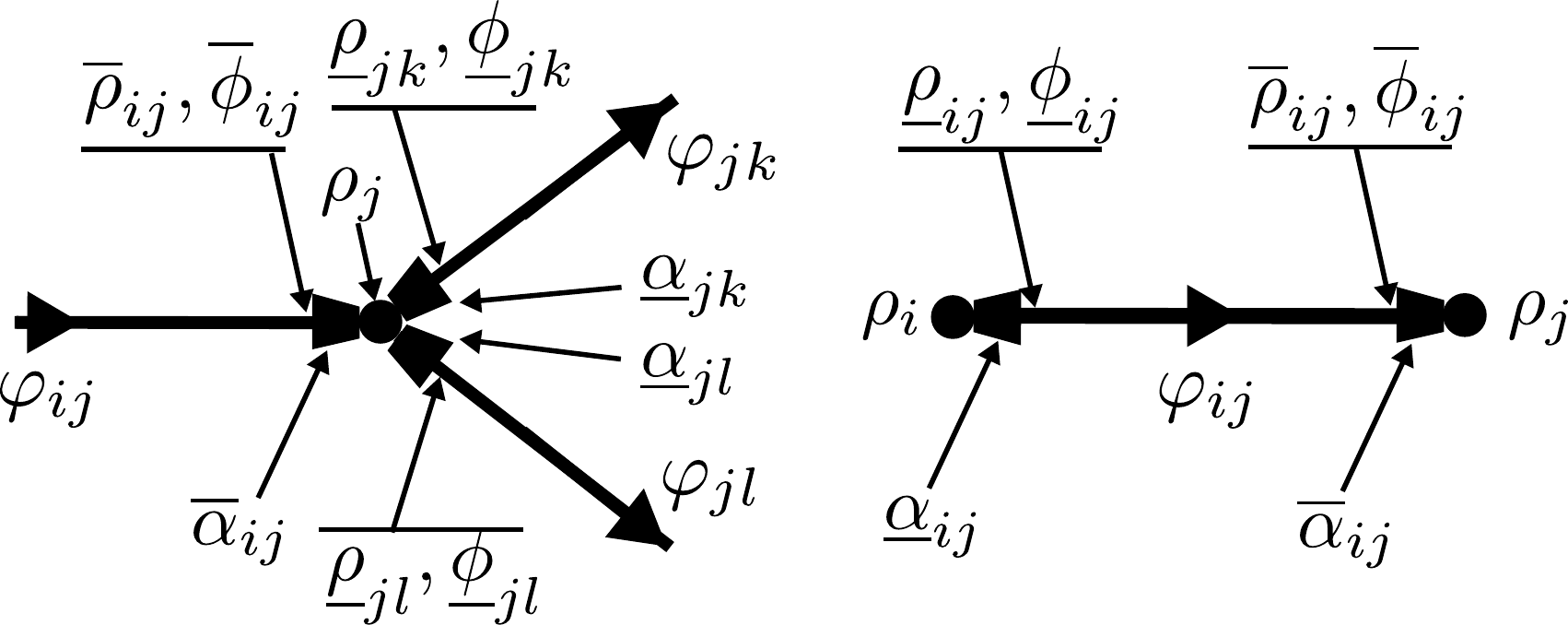} \vspace{-2ex} 
\caption{\textbf{Illustration of pipeline network modeling notation.}  Nodal densities  $\rho_j$, edge endpoint variables $\underline{\rho}_{ij}$, $\underline{\phi}_{ij}$, $\overline{\rho}_{ij}$, and $\overline{\phi}_{ij}$, and actuators $\underline{\alpha}_{ij}$, $\overline{\alpha}_{ij}$ are labeled for an edge (left) and for a joint (right).} \label{fig:schematic}
\vspace{-2ex} 
\end{figure}
Borrowing from power systems nomenclature, we distinguish between two different types of nodes. The set of ``slack'' nodes, denoted by $\mathcal V_\sigma \subset \mathcal V$, are nodes where the mass flow rate is controlled to achieve a prescribed (time-varying) density $\sigma_j(t)$,  
\begin{equation}
\rho_{i}(t)=\sigma_i(t). \,\forall\, i\in\mathcal V_\sigma. \label{eq:slack_pressure}
\end{equation}
The set of ``non-slack'' nodes corresponds to the remaining nodes $j \in \mathcal V_d = \mathcal V \setminus \mathcal V_\sigma$, where the mass flow $d_j(t)$ is the prescribed control variable. While this distinction is less important for optimization problems, it matters for simulations.

\subsubsection{Gas Compressor Modeling} 
Gas compressors are used to control the flow of gas and maintain pressure throughout a pipeline. While each compressor station often contains multiple compressor machines, we will refer to entire stations as compressors.  
In our model, each compressor is located at a node, and controls the change of gas density between the node (inlet pressure) and the beginning of the pipeline it is connected to (outlet pressure). While some compressors utilize gas from the pipeline for their own operation, we assume that the mass flow between inlet and outlet is conserved.

Each compressor is indexed by $(i,j,\chi)$ in the set of compressors $\mathcal C\subset \mathcal E\times\{+,-\}$, where $\chi\in\{+,-\}$ indicates orientation. The compressor $(i,j,+)\in\mathcal{C}$ is located at node $i\in\mathcal V$ and adjusts density of gas flowing into edge $(i,j)\in\mathcal E$ in the $i \to j$ direction. The compressor $(i,j,-)\in\mathcal{C}$ is located at node $j\in\mathcal V$ and adjusts density into edge $(i,j)\in\mathcal E$ in the direction $j\to i$. The amount of compression is given as a multiplicative ratio $\underline{\alpha}_{ij}\geq 1$ for $\forall (i,j,+)\in\mathcal{C}$ and $\overline{\alpha}_{ij}\geq 1$ for $\forall (i,j,-)\in\overline{\mathcal{C}}$.  
This leads to relationships between inlet and outlet pressures,
\begin{subequations} \label{eq:density_bal}
\begin{flalign}
& \underline{\rho}_{ij}(t) = \underline{\alpha}_{ij}(t)\rho_i(t), \, \forall \, (i,j,+) \in \mathcal{C}, & \label{eq:nodal_density_balance_1}\\
& \overline{\rho}_{ij}(t) = \overline{\alpha}_{ij}(t)\rho_j(t), \, \forall \, (i,j,-) \in \mathcal{C}, & \label{eq:nodal_density_balance_2}
\end{flalign}
\end{subequations}

\subsection{Operational constraints}

\subsubsection{Pressure constraints} %
Too high pressure can lead to damage of equipment, while too low pressure \avz{may violate delivery contracts.} It is therefore necessary to enforce upper and lower pressure bounds $\rho^{max}, \rho^{min}$ throughout the system.
For the pipes, it is sufficient to enforce constraints only at the endpoints, because friction effects result in monotone decrease of pressure along the direction of flow \cite{zlotnik16ecc}. %
We express these constraints as
\begin{subequations} \label{eq:density_limits}
\begin{align}
{\rho}_{ij}^{\min} \leq \underline{\rho}_{ij}(t)  &\leq {\rho}_{ij}^{\max}, \quad \fA~ (i,j)\!\in\!\cE \label{eq:density_in} \\
{\rho}_{ij}^{\min} \leq \bar{\rho}_{ij}(t)  &\leq {\rho}_{ij}^{\max}, \quad \fA~ (i,j)\!\in\!\cE \label{eq:density_out} \\
\rho_i^{\min}\leq \rho_i(t) &\leq \rho_i^{\max},  \quad \fA~ i\!\in\!\cV. \label{eq:density_min}
\end{align}
\end{subequations}
\subsubsection{Compressor constraints} The compression schedule is constrained by the following inequalities,
\begin{subequations} \label{eq:compressor_limits}
\begin{align}
\underline{\varepsilon}_{ij}|\underline{\phi}_{ij}(t)| ((\underline{\alpha}_{ij}(t))^{h_g}-1) \leq \underline{E}_{ij}^{\max}, \,\,\, \fA~(i,j,+)\in \mathcal C, \label{eq:comppow1a}\\
\overline{\varepsilon}_{ij}|\overline{\phi}_{ij}(t)| ((\overline{\alpha}_{ij}(t))^{h_g}-1) \leq \overline{E}_{ij}^{\max}, \,\,\, \fA~(i,j,-)\in  \mathcal C, \label{eq:comppow1b} \\
\underline{\alpha}_{ij}(t) \geq 1, \quad \overline{\alpha}_{ij}(t) \geq 1, \quad \fA~ (i,j,\chi)\in\mathcal C. \label{eq:compmin}
\end{align}
\end{subequations}
Here \eqref{eq:comppow1a}-\eqref{eq:comppow1b} limit the energy (or power) used by compressors, and \eqref{eq:compmin} reflects that they are designed and operated only to boost pressure.  Here $h_g=(\gamma-1)/\gamma<1$, where $\gamma$ is the specific heat capacity ratio of the gas, and $\underline{\varepsilon}_{ij}$ and $\overline{\varepsilon}_{ij}$ correspond to $\varepsilon=\eta \cdot T_1 / (e_a \cdot e_m \cdot G \cdot h_g)$ for $(i,j,+)$ and $(i,j,-)$, respectively, where the discharge temperature, adiabatic and mechanical efficiencies, gas gravity, and conversion factor are represented by $T_1$, $e_a$, $e_m$, $G$, and $\eta$, respectively \cite{menon05}.

\subsubsection{Gas injection constraints} The withdrawal $d_j(t)$ and supply $s_j(t)$ of natural gas at each node may be optimized and constrained according to 
\begin{subequations} \label{eq:withdrawal_limits}
\begin{align}
0 \leq d_j^{\min}(t) \leq d_j(t) \leq d_j^{\max}(t), & \text{ if $j$ is a consumer,} \label{eq:demand_limits} \\ s_j^{\min}(t) \leq s_j(t) \leq s_j^{\max}(t) \leq 0, & \text{ if $j$ is a supplier.} \label{eq:supply_limits}
\end{align}
\end{subequations}
We may set $d_j^{\min}(t) = d_j^{\max}(t)$ and $s_j^{\min}(t) = s_j^{\max}(t)$ for all $t\in\mathcal [0,T]$ to specify a given (non-optimized) withdrawal or supply profile at a node $j\in\cV$.

\subsubsection{Time periodicity constraints} In previous studies \cite{zlotnik15cdc,zlotnik17tpwrs,zlotnik19cdc}, we found that the initial and terminal conditions must satisfy time-periodicity in order to generate computationally well-posed problems. This formulation requires
\begin{subequations}
\begin{flalign}
&\rho_{ij}(0,x_{ij})=\rho_{ij}(T,x_{ij}),  \,\forall\, (i,j)\in\mathcal E, &\label{eq:termcon0a}\\
&\phi_{ij}(0,x_{ij})=\phi_{ij}(T,x_{ij}),  \,\forall\, (i,j)\in\mathcal E. &\label{eq:termcon0b}
\end{flalign}
\label{eq:termcon0}
\end{subequations}
\noindent Because of the nodal compatibility conditions, this implies that the compression schedule and gas withdrawals need to satisfy the same conditions:
\begin{subequations}
\begin{flalign}
&\underline{\alpha}_{ij}(0)=\underline{\alpha}_{ij}(T), \, \overline{\alpha}_{ij}(0)=\overline{\alpha}_{ij}(T),  \,\forall\, (i,j,\chi)\in\mathcal C, &\label{eq:termcon1c} \\
&d_j(0)= d_j(T), \,\forall\, j \in \mathcal V_d. &\label{eq:termcon1d}
\end{flalign}
\label{eq:termcon1}
\end{subequations}
\noindent We claim that for the above regularity conditions to be satisfied, it is sufficient that the boundary conditions and constraint bound values are time-periodic.  That is,
\begin{subequations}
\begin{align}
&d_j^{\min}(0) = d_j^{\min}(T), \, d_j^{\max}(0)=d_j^{\max}(T),  \,\forall\, j\in \mathcal V_d, &\label{eq:termcon2a} \\
&s_j^{\min}(0) = s_j^{\min}(T), \, s_j^{\max}(0)=s_j^{\max}(T),  \,\forall\, j\in \mathcal V_d, &\label{eq:termcon2b} \\
&\sigma_j(0)=\sigma_j(T), \,\forall\, j \in \mathcal V_\sigma. &\label{eq:termcon2c}
\end{align}
\label{eq:termcon2}
\end{subequations}
\edit{\noindent We argue that if all parameter functions are time-periodic, the geometry of the optimal solution will be toroidal in the space-time manifold, and thus time-periodic as well.}

\subsection{Optimal Transient Gas Flow Scheduling} \label{subsec:optflowschedule}
Here we formulate an optimal control problem for scheduling natural gas flow over a pipeline network, \edit{which reflects the business goals and operational requirements of gas pipeline system managers \cite{zlotnik2019pipeline,zlotnik19cdc}.}
The decision variables are time-varying compressor ratios $\underline{\alpha}_{ij}$ for $(i,j,+)\in \mathcal C$ and $\overline{\alpha}_{ij}$ for $(i,j,-)\in \mathcal C$, and the nodal gas withdrawals $d_j$ for $j\in\mathcal V_d$. 
We suppose that slack node densities $\sigma_j(t)$ are given, while other densities $\rho_j(t)$ and the flows $\phi_{ij}(t)$ are considered dependent state variables that are determined as a function of the decision variables. 
The optimization problem maximizes the economic value provided by the pipeline for its users, while minimizing the cost of operation over a time horizon $\mathcal T=[0,T]$. %

\subsubsection{Decision Variables and Parameters}
We denote decision variables for pipeline system scheduling with the shorthand
\begin{subequations}    \label{eq:decision_variables_gas}
\begin{align}
     &\mbox{Compressor ratios:} \,\, &&\boldsymbol{\alpha} \equiv (\underline{\alpha}_{ij}(t), \overline{\alpha}_{ij}(t), \, (i,j,\chi)\in\mathcal{C}),  \nonumber\\
    &\mbox{Gas withdrawals:} \, &&\mathbf{d} \equiv (d_j(t), \,\, j\in\mathcal V ),   \nonumber\\
    &\mbox{Gas supply:} \, &&\mathbf{s} \equiv (s_j(t), \,\, j\in\mathcal V ),   \nonumber\\ %
    &\mbox{Densities:} \, &&\boldsymbol{\rho} \equiv (\rho_j(t), \,\, j\in\mathcal V ),   \nonumber\\ %
    &\mbox{Flows:} \, &&\boldsymbol{\phi} \equiv (\bar{\phi}_{ij}(t), \underline{\phi}_{ij}(t), \,\, (i,j)\in\mathcal{E}),  \nonumber %
\end{align}
\end{subequations}

\subsubsection{Deterministic Gas Flow Constraints}
We collect the above equality and inequality equations that specify physics and engineering constraints for a gas pipeline system into the constraint set $\Gamma$:
\begin{subequations}    \label{eq:gas_feasibility_set}
\begin{align}
&\mathbf{\Gamma}(\boldsymbol{\alpha},\mathbf{d},\mathbf{s},\boldsymbol{\rho},\boldsymbol{\phi})  = \nonumber\\
&\quad \begin{cases}
& \mbox{Gas flow dynamic equations}~\eqref{eq:pde_1}, \nonumber \\
& \mbox{Mass flow balance}~\eqref{eq:massflowbal1}, \nonumber \\
& \mbox{Slack node density}~\eqref{eq:slack_pressure}, \nonumber \\
& \mbox{Compressor action}~\eqref{eq:density_bal}, \nonumber\\
& \mbox{Density limits}~\eqref{eq:density_limits},  \nonumber\\
& \mbox{Compressor limits}~\eqref{eq:compressor_limits},  \nonumber\\
& \mbox{Withdrawal limits}~\eqref{eq:withdrawal_limits},  \nonumber \\
& \mbox{Time periodicity}~\eqref{eq:termcon0}-\eqref{eq:termcon2}. \nonumber 
\end{cases}
\end{align}
\end{subequations}

\subsubsection{Objective Functions}
\edit{The business goal of pipeline managers is to maximize profit, leading to the objective function}
\begin{align} \label{eq:obj_profit}
J_E(\mathbf{d})\ & \triangleq  \sum_{j\in\mathcal{V}} \int_0^T (c_{d,j}(t)d_j(t)+c_{s,j}(t)s_j(t)) dt.
\end{align}
Here, $c_{d,j}(t)\geq 0$ is the bid (or offer price) of the consumer $d_j(t)$, and $c_{s,j}(t)\leq 0$ is the bid of the supplier $s_j(t)$.

To promote the efficient compression operation, we also minimize the power used by compressors, which is given by
\begin{align} \label{eq:obj_eff}
J_C(\boldsymbol{\alpha}) & \triangleq  \sum_{(i,j)\in\mathcal{C}} \int_0^T \underline{\varepsilon}_{ij}|\underline{\phi}_{ij}(t)| ((\underline{\alpha}_{ij}(t))^{h_g}-1) dt.
\end{align}

\subsubsection{Optimal Pipeline Flow Problem}
With the above engineering and physical constraints, the deterministic optimal control problem is
\begin{subequations} \label{eq:dogf}
\begin{align}
    &\!\!\!\!\!\max\limits_{\boldsymbol{\alpha},\mathbf{d},\mathbf{s},\boldsymbol{\rho},\boldsymbol{\phi}}&& \mbox{Profit \& Efficiency: $J_E(\mathbf{d})-J_C(\boldsymbol{\alpha})$} \\
&\mbox{s.t. \,\,\,}
&& \mbox{Gas flow constraints:} ~\mathbf{\Gamma}(\boldsymbol{\alpha},\mathbf{d},\mathbf{s},\boldsymbol{\rho},\boldsymbol{\phi}).
\end{align}
\end{subequations}

\subsection{Robust Optimal Gas Flow Scheduling}
\label{subsec:robust_gas}
\avz{In gas pipelines, uncertainty arises in gas withdrawals $\mathbf{d}$, as well as in the state of the system (i.e, uncertainty about the exact density state at the beginning of the day).} In this section, we review a key monotonicity property of natural gas flow, and how these results form the necessary theoretical foundation for extending the deterministic optimization problem \eqref{eq:dogf} to provide robust feasibility guarantees for a range of gas withdrawals. %

\subsubsection{Monotonicity Property of Gas Pipeline Flows}
Our approach to ensure feasibility is inspired by approaches that examine stability and robustness of distributed routing solutions \cite{como13a,como13b}.  These studies utilize monotone operator theory to demonstrate that the dynamics in question are monotone control systems \cite{angeli03,como10,lovisari14,sootla2018operator}.  Such systems, typically studied in the context of ordinary differential equations (ODEs), possess a monotone order propagation property with respect to certain input variables  \cite{kamke32,hirsch85,smith88,hirsch05}.  The concept of monotone control systems can enable robust optimal control of very complex networked energy \avz{systems} \cite{angeli03,angeli04}, including dynamic natural gas pipelines \cite{vuffray15cdc,zlotnik16ecc,misra2016monotone}.

\emph{The monotonicity results for gas pipeline flows show that pressure anywhere in the network can only increase monotonically when more gas is injected anywhere in the system.} Conversely, the pressure can only decrease if more gas is withdrawn from the system.  While intuitive for a single pipe, this property is not easy to prove for general systems with loops. The Aquarius theorem \cite{vuffray15cdc} establishes conditions under which monotonicity holds for steady-state natural gas flows. This result has been extended to the time-varying setting with transient gas flows and compressor behavior in \cite{zlotnik16ecc,misra2016monotone}. We summarize one version of this result for completeness.

Let us define a shorthand for the density state of a pipeline network by $\boldsymbol{\rho}(t)\equiv (\rho_{i}(t), \,\, i\in\mathcal V)$, and assume that two initial density states $\boldsymbol{\rho}^{(1)}(0)$ and $\boldsymbol{\rho}^{(2)}(0)$ are given for a pipeline system. Here, $\boldsymbol{\rho}^{(1)}(0) \leq \boldsymbol{\rho}^{(2)}(0)$, i.e. the densities in the state $\boldsymbol{\rho}^{(1)}(0)$ are lower than those in state $\boldsymbol{\rho}^{(2)}(0)$ pointwise everywhere in the network.  Starting from this initial condition, the lower density state $\boldsymbol{\rho}^{(1)}(0)$ is subjected to withdrawal profile  $\mathbf{d}^{(1)}(t)$ and the higher density state is subjected to withdrawal profile $\mathbf{d}^{(2)}(t)$. The widthdrawal at any point in the network is greater for $\mathbf{d}^{(1)}(t)$ than for $\mathbf{d}^{(2)}(t)$ (conversely, any injection for $\mathbf{d}^{(1)}(t)$ is less than for $\mathbf{d}^{(2)}(t)$), such that $\mathbf{d}^{(1)}(t) \geq \mathbf{d}^{(2)}(t)$ for all $t \geq 0$. If, in addition, the compressor ratio protocols $\boldsymbol{\alpha}$ are kept fixed among all scenarios\footnote{This assumption can be relaxed under certain conditions, see \cite{zlotnik16ecc, misra2016monotone}.}, the monotonicity theorem states that the densities in scenario (1) will remain lower than the densities in scenario (2) for all times, i.e., $\boldsymbol{\rho}^{(1)}(t) \leq \boldsymbol{\rho}^{(2)}(t)$ for all $t \geq 0$. 
We refer the reader to additional sources on the monotonicity property \cite{zlotnik16ecc, misra2016monotone,misra20pieee}.

\subsubsection{Robust Optimal Gas Flow Scheduling}
We now describe how the monotonicity properties can be utilized to ensure robustly feasible gas \avz{pipeline} operation. 
Consider a pipeline system that, starting from an initial state $\boldsymbol{\rho}(0)$, is subject to an uncertain, but bounded gas withdrawal profile $\mathbf{d}(t)$, %
\begin{equation}
    \overline{\mathbf{d}}^{(2)}(t) \leq \mathbf{d}(t) \leq \overline{\mathbf{d}}^{(1)}(t).
    \label{eq:withdrawal_bounds}
\end{equation}
We wish to extend the formulation \eqref{eq:dogf} to ensure that the solution will be feasible for any withdrawal profile $\mathbf{d}(t)$ within the considered range. This robust optimal control problem is a semi-infinite program, where the solutions must be robust to a continuum of possible withdrawal trajectories (i.e., an infinite number of constraints) with corresponding compressor ratios and supply injections (i.e., an infinite number of decision functions). 
The monotonicity property enables us to reduce this to a finite-dimensional optimal control problem, where the constraints are enforced only for the extreme scenarios $\overline{\mathbf{d}}^{(1)}(t)$ and $\overline{\mathbf{d}}^{(2)}(t)$, and a single operating schedule $\boldsymbol{\alpha}(t)$ and $\mathbf{s}(t)$ is shared among all the withdrawal profiles. We suppose that a single operating schedule for the compressors and controllable injections is of practical interest, as it is challenging for the pipeline system operator to assess the future (uncertain) trajectory of the system and change the controls in real time. 
The robust optimal gas flow scheduling problem is given by
\begin{subequations} \label{eq:robust_dogf}
\begin{align}
    &\!\!\!\!\!\max\limits_{\boldsymbol{\alpha},\mathbf{d},\boldsymbol{\rho},\boldsymbol{\phi}}&& \mbox{Profit \& Efficiency: $J_E(\mathbf{d})-J_C(\boldsymbol{\alpha})$} \\
&\mbox{s.t.: \,\,\,}
&& \mbox{High gas withdrawals:} ~\mathbf{\Gamma}(\boldsymbol{\alpha},\mathbf{d}^{(1)},\mathbf{s},\boldsymbol{\rho},\boldsymbol{\phi}) \nonumber \\
&&& \mbox{Low gas withdrawals:} ~\mathbf{\Gamma}(\boldsymbol{\alpha},\mathbf{d}^{(2)},\mathbf{s},\boldsymbol{\rho},\boldsymbol{\phi}) \nonumber \\
&&& \mbox{Withdrawal bounds}~\eqref{eq:withdrawal_bounds}\nonumber
\end{align} 
\end{subequations}
The monotonicity property of the gas flows hence provides robust feasibility guarantees by simply doubling the number of constraints, enabling tractable robust optimization.

\section{Integrated Electricity and Gas Network Uncertainty Management Frameworks} \label{sec:integration}

In this section, we model the interactions between the electric power and natural gas systems, and \edit{combine the above models for independent electricity and natural gas flow optimization under uncertainty into a new optimization model} for integrated scheduling and uncertainty management.

\subsection{Coupling of Electricity and Gas Delivery Networks}
The electric grid and natural gas pipeline system are coupled primarily through gas-fired power plants, whose electric power production determines the demand for gas. While there are other interactions, such as the electricity consumption of natural gas equipment (e.g., the electric power consumption by pipeline compressor stations) or emerging ``power-to-gas'' (P2G) technology, the coupling through gas fired generators is by far the dominant contributing factor today and is the sole focus of this paper.

The real-time gas-fired electricity generation is determined by the scheduled generation $\generationsetpoint(t)$ and the activation of reserves, which is bounded by the reserve allocation. This yields bounds on generation by gas fired power plants, of form 
\begin{align}
& \mathbf{p}^{\min}(t) \leq \generationsetpoint(t) - \beta(t)\Omega(t) \leq \mathbf{p}^{\max}(t), \quad    \nonumber\\[2pt]
& \text{with}~\mathbf{p}^{\max}(t) = \generationsetpoint(t) + r^{+}(t) ~\text{and}~
 \mathbf{p}^{\min}(t) = \generationsetpoint(t) - r^{-}(t). \nonumber 
\end{align}
The relationship between real time electricity production and nominal gas consumption of a generator is captured by its heat rate curve $q_{\mathrm{heat}}$, which we approximate as a linear function,
\begin{align}   \label{eq:coupling_realtime}
    \tilde{d}(t) = q_{\mathrm{heat}}(\tilde{p}(t)) = c_0 + c_1 (\generationsetpoint(t) - \beta(t)\Omega(t)). 
\end{align}
We use the scheduled power generation and reserve capacities to construct three ordered demand profiles for natural gas: 
\begin{subequations}
\label{eq:coupling_robust}
(0) A \emph{nominal} profile $\mathbf{d}^{\mathrm{nom}}(t)$ corresponding to the scheduled  generation setpoint $\mathbf{\generationsetpoint}(t)$
\begin{equation}
    \mathbf{d}^{\mathrm{nom}}(t) = q_{\mathrm{heat}}(\mathbf{\generationsetpoint}(t))
    \label{eq:coupling_nominal}
\end{equation}
(1) A \emph{high-demand} profile $\mathbf{d}^{\max}(t)$ corresponding to full activation of upwards reserves $\mathbf{p}^{\max}(t)$,
\begin{equation}
    \mathbf{d}^{\max}(t) = q_{\mathrm{heat}}(\mathbf{\generationsetpoint}(t) + r^{+}(t))
\end{equation}
(2) A \emph{low-demand} profile $\mathbf{d}^{\min}(t)$ corresponding to full activation of downwards reserves $\mathbf{p}^{\min}(t)$,
\avz{
\begin{equation}
    \mathbf{d}^{\min}(t) = q_{\mathrm{heat}}(\mathbf{\generationsetpoint}(t) - r^{-}(t))
\end{equation}
}
\end{subequations}
\avz{Approximating the heat rate curve \eqref{eq:coupling_realtime} as monotonically increasing with} power production, the nominal fuel demand of gas-fired generators will satisfy $\mathbf{d}^{\min}(t) \leq \tilde{d}(t) \leq \mathbf{d}^{\max}(t)$.
The above bounds on generator gas demand enable seamless employment of the monotonicity-based robust gas network feasibility described in Section~\ref{sec:gas}.

\subsection{Proposed Method for Integrated Uncertainty Management}
Given the above modeling considerations, we propose a method for integrated scheduling and uncertainty management of electricity and gas delivery networks. Our formulation computes an optimal generation schedule and reserve for a power grid subject to fuel constraints that are derived from the gas pipeline system. The formulation considers the impact of renewable energy fluctuations in the power grid, which translates into uncertain, but bounded gas withdrawals as described by \eqref{eq:coupling_robust}. 
This is achieved by combining the chance-constrained formulation for the electric grid (which guarantees feasibility of electric grid operation under uncertain renewable energy generation with high probability, as long as sufficient gas is available) with the robust formulation for the natural gas network (which guarantees that the natural gas network can provide sufficient delivery to support both the nominal generation schedule and reserves). The formulation is given by
\begin{subequations} \label{eq:gas_grid_robust}
\begin{align}
    &\min\limits_{\substack{\mathbf{\generationsetpoint},\mathbf{r},\beta,\\\boldsymbol{\alpha},\mathbf{s},  \mathbf{d}^{\max},\\\mathbf{d}^{\textrm{nom}},\mathbf{d}^{\min}}} &&  \mbox{Joint obj.: $J_P(\mathbf{\generationsetpoint},\mathbf{r})+J_C(\boldsymbol{\alpha})$} \\
&\mbox{s.t.: \,\,\,}
&& 
\mbox{Power scheduling:}~\Pi(\mathbf{\generationsetpoint},\mathbf{r},\boldsymbol{\beta}) \\
&&& \mbox{Coupling constraints:}~\eqref{eq:coupling_robust} \\
&&& \mbox{High withdrawals:}~\mathbf{\Gamma}(\boldsymbol{\alpha},\mathbf{d}^{\max},\mathbf{s},\boldsymbol{\rho},\boldsymbol{\phi}) \\
&&& \mbox{Nominal withdrawals:}~\mathbf{\Gamma}(\boldsymbol{\alpha},\mathbf{d}^{\textrm{nom}},\mathbf{s},\boldsymbol{\rho},\boldsymbol{\phi}) \\
&&& \mbox{Low withdrawals:}~\mathbf{\Gamma}(\boldsymbol{\alpha},\mathbf{d}^{\min},\mathbf{s},\boldsymbol{\rho},\boldsymbol{\phi})
\end{align}
\end{subequations}
We note that the gas withdrawal profiles for all customers except the gas-fired power plants are assumed to be known and pre-determined. However, the formulation could easily be extended to account for uncertainty in those load profiles, or to include them as load profiles that are optimized.
\edit{Further, we note that the objective function no longer includes the profit maximization term $J_E(\mathbf{d})$ for the natural gas system. \avz{We suppose that in today's intra-day energy markets, a natural gas pipeline achieves most profits by selling gas transportation to gas-fired generators. Thus, it is in the interest of the pipeline manager to be able to supply the gas demanded by the electric market clearing outcome. The compressor energy minimization objective $J_C(\boldsymbol{\alpha})$ is included to regularize the solution and break any degeneracy.} In other situations, e.g. to optimize the transport of gas to customers in addition to gas-fired generators, it would be natural to also include the profit term $J_E(\mathbf{d})$ in the cost function in the integrated problem. Such a change to the objective function could be easily made. }

\section{Case Study: Analysis Framework}
\label{sec:computation}

In this section, we describe the assessment framework we use to evaluate performance and quantify the effectiveness of our proposed approach, using a computational study based on a coupled power system and gas pipeline test case. We describe the test case, \avz{the full formulation, and two intermediate benchmark formulations for comparison}, as well as the evaluation workflow below. The test results are discussed in the following section.

\subsection{Benchmark Formulations}
The intention of this comparison is to evaluate the consideration of robustness in the joint optimization of electric grids and natural gas networks. Both our proposed formulation and the two benchmark formulations are therefore joint optimization problems that consider full information sharing among the two systems, but include different levels of consideration regarding uncertainty management and feasibility of reserve provision. 

\noindent \textbf{Formulation 0:} \textit{\textbf{Deterministic Power + Deterministic Gas: }} 
This benchmark formulation is a deterministic scheduling problem, which assumes $\omega(t)=0$ for the power grid scheduling and instead includes pre-determined (fixed) reserve capacities $\mathbf{r}$ and participation factors $\boldsymbol{\beta}$. The feasibility of the natural gas network only considers the nominal generation schedule, and does not account for the fixed values of $\mathbf{r}$. This gives rise to the following formulation:
\begin{subequations} \label{eq:gas_grid_deterministic}
\begin{align}
    \min\limits_{\mathbf{\generationsetpoint,\alpha,d^{\textrm{nom}},s}}& \  \mbox{Joint obj.:
    $J_P(\mathbf{\generationsetpoint})+J_C(\boldsymbol{\alpha})$} \\
\mbox{s.t.: \,\,\,}
& \mbox{Power flow constraints:} ~\Pi(\mathbf{\generationsetpoint}) \\
& \mbox{Coupling constraints:} ~\mbox{Eq.}~\eqref{eq:coupling_nominal} \\
&\mbox{Gas flow constraints:}~\mathbf{\Gamma}(\boldsymbol{\alpha},\mathbf{d}^{\textrm{nom}},\mathbf{s},\boldsymbol{\rho},\boldsymbol{\phi})
\end{align}
\end{subequations}

\noindent \textbf{Formulation 1:} \textit{\textbf{Chance-Constrained Power + Deterministic Gas: }} 
This benchmark formulation includes chance constraints to account for uncertainty in power system operation scheduling, but takes only the nominal generation schedule into account when checking feasibility and forming a compression schedule in the natural gas network. The effect of reserve procurement on real-time fluctuations in gas demand is therefore not considered. %
\begin{subequations} \label{eq:gas_grid_nominal}
\begin{align}
    \min\limits_{\mathbf{\generationsetpoint,\beta,r,\alpha,d^{\textrm{nom}},s}}& \  \mbox{Joint obj.:
    $J_P(\mathbf{\generationsetpoint},\mathbf{r})+J_C(\boldsymbol{\alpha})$} \\
\mbox{s.t.: \,\,\,}
& \mbox{Power flow constraints:} ~\Pi(\mathbf{\generationsetpoint},\mathbf{r},\boldsymbol{\beta}) \\
& \mbox{Coupling constraints:} ~\mbox{Eq.}~\eqref{eq:coupling_nominal} \\
& \mbox{Gas flow constraints:}~\mathbf{\Gamma}(\boldsymbol{\alpha},\mathbf{d}^{\textrm{nom}},\mathbf{s},\boldsymbol{\rho},\boldsymbol{\phi})
\end{align}
\end{subequations}

\begin{figure*}[!t]
\centering
\includegraphics[width=.9\textwidth]{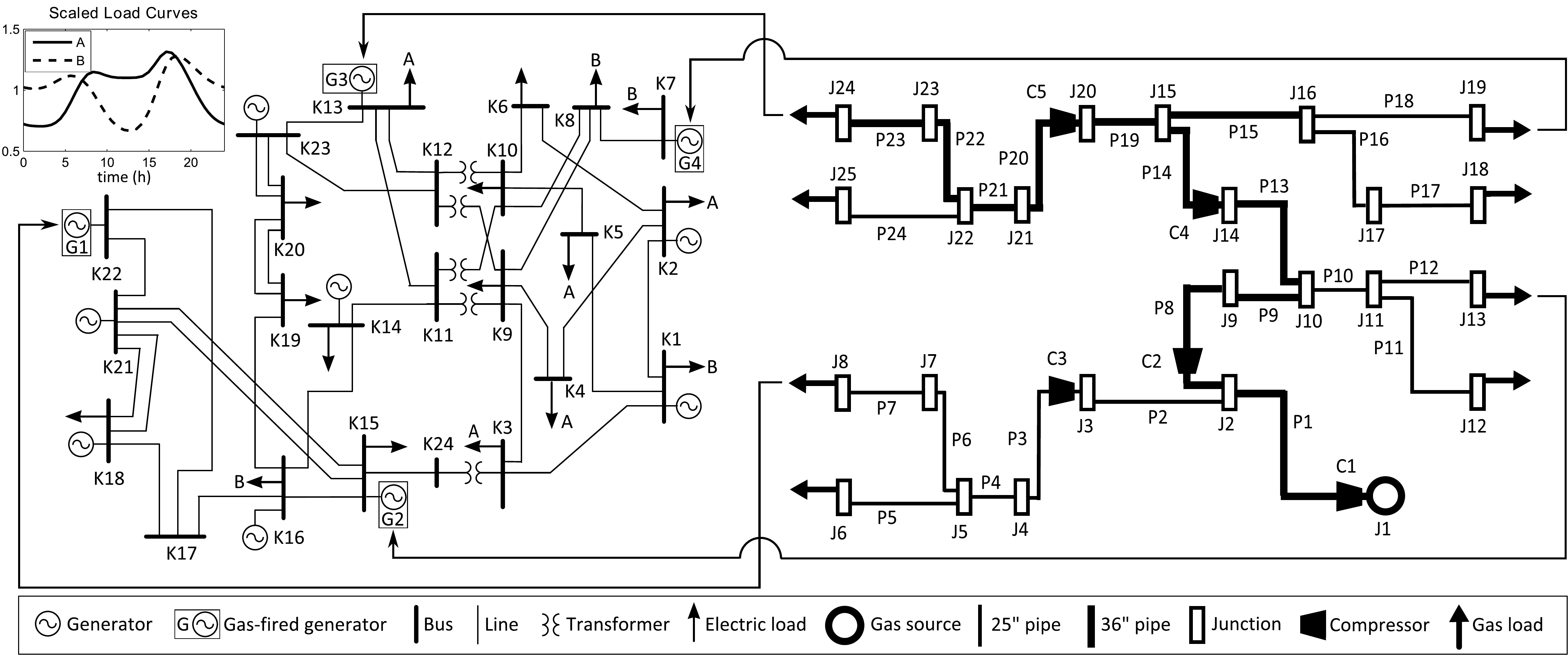}
\caption{\textbf{Schematic of integrated electric and gas test networks (reproduced from \cite{zlotnik17tpwrs}).}  IEEE RTS96 One Area 24 node power system (left) coupled to the 24 pipe benchmark gas pipeline (right) through gas-fired generators (G1 to G4).  Electric loads marked with A or B are scaled by the time-varying load curves in the inset plot at top left.  Electric buses (K1 to K24), gas pipes (P1 to P24), pipe junctions (J1 to J25), and gas compressors (C1 to C5) are indicated.}
\label{fig:gas_grid_test_system}
\vspace{-2ex}
\end{figure*}

\noindent \textbf{Formulation 2:} \textit{\textbf{Chance-Constrained Power + Robust Gas: }} 
Our last formulation is the proposed method \eqref{eq:gas_grid_robust}, which includes chance constraints to account for uncertainty in power system operation scheduling, as well as the three scenarios required to guarantee feasibility of the gas pipeline system.

\subsection{Test Case and Implementation}
We synthesize a case study based on the data from our previous work \cite{zlotnik17tpwrs}, \cite{gasgridgithub}. 
We investigate different loading conditions, with mean electricity demand equal to 90\% of the maximum load level, scaled using the profiles in Figure \ref{fig:gas_grid_test_system}. The uncertainty in the electric power demand $\omega$ is modeled as a multivariate normal distribution with zero mean and variance equal to 3.5\% of the load, i.e., $0.035\cdot h(t)$. The acceptable violation probabilities are set to $\varepsilon = 0.1$ for the line constraints and $\varepsilon=0.01$ for the generator constraints. 
The ramp rates for the generators are adapted from \avz{a previous study} \cite{barrows2019}.

In the gas network, the demand profiles for the gas-fired generators are derived from the scheduled power generation and the reserve activation, as described in Section \ref{sec:integration}. All other \avz{gas pipeline loads} are set to constant values. There are four nodes with \avz{gas-fired} generators G1-G4, as seen in Figure \ref{fig:gas_grid_test_system}. Each of those generation nodes includes several generation units. The heat rate parameters in \eqref{eq:coupling_realtime} with $c_0=0$, and $c_1=10$ mmbtu/MWh for the peaking plant (6 generation units in G1 and 5 generation units in G2), and $c_1=15$ mmbtu/MWh for combined cycle plants (1 generation \avz{unit} in G2, and 3 generation units in both G3 and G4). %

In alignment with standard power system \avz{operating} practices, we solve the system over a time~horizon of 24 hours. However, because the intra-day generation dispatch is not necessarily time periodic (particularly not when considering the activation of \avz{reserves}), we extend the time~horizon of the computation to 30 hours and enforce that the system is time-periodic over this extended horizon. This allows us to compute a schedule for the non-time-periodic trajectories. The objective function only considers the first 24 hours.
While the original test system only includes loading scenarios for each hour, we need smaller time discretization to accurately capture the transient dynamics of the gas flow. For the optimization problem, we extend the data set to have time steps of 30 minutes, while the simulations assume 10 minute time steps. The load at each time step is obtained by interpolating between the load at each hour.

\subsection{Framework for Evaluation}
We evaluate the performance of our method using the following workflow.
\begin{enumerate}
    \item Choose load and uncertainty levels for the test case.
    \item Compute schedule solutions to Formulations 0, 1 and 2 to obtain a schedule for the power and gas networks.
    \item Run a Monte Carlo simulation to assess the system performance.
\end{enumerate}
We elaborate more on the Monte Carlo simulation and evaluation metrics that we use to assess the performance.

\subsubsection{Monte Carlo simulation} 
To obtain statistics of system performance, such as constraint violations, we run Monte Carlo simulations. As previous  studies have investigated the impact of renewable energy fluctuations on power system constraint violations in detail \cite{roald2013analytical,bienstock2014chance,roald2016corrective}, the present study focuses on constraint violations in the natural gas system and their impact in terms of load shed by the power system.
In each simulation, we generate a time-series for the uncertain electric load time series by drawing a new realization of the deviation $\omega$ every 10 minutes.
We then solve the DC power flow \eqref{eq:ptdf} for each hour, where we assume that the generators are adjusted to balance load according to the reserve activation policy \eqref{eq:gen_policy}. 
The generation by gas-fired power plants is translated to a real-time gas withdrawal $\mathbf{\tilde{d}}^{\mathrm{nom}}$ using the generator heat rate \eqref{eq:coupling_realtime}. 
We then formulate an initial boundary value problem (IBVP) for the natural gas network. As initial conditions, we utilize the first timestep from the nominal gas flow trajectory obtained from the optimization problem solution.
The compression ratios $\boldsymbol{\alpha}$ are also obtained from the optimization problem solution. 
Given these parameters, the IBVP has a unique solution obtained through a simulation using a DAE discretization \cite{sundar18tcst}.  

\subsubsection{Pressure Bound Violations}
The IBVP simulation provides us with the density trajectories for each node in the pipeline network. Based on these trajectories, we calculate the maximum pressure constraint violation,
\begin{align} \label{eq:pviol_max}
\!\!\!\! V_{\mathbf{p}}^{\max} \! = \! \max_t(\pi^{\min}-\pi(t))_+
\end{align}
and the maximum pressure constraint violation $L_2$  norm
\begin{align} \label{eq:pviol_norm}
||V_{\mathbf{p}}||_2 \! = \! \frac{1}{N_h}\sum_{t\in\mathcal{T}}\sum_{i\in\mathcal{V}}(\pi^{\min}-\pi(t))_+^2 %
\end{align}
where $(x)_+=x$ if $x\geq 0$ and $(x)_+\equiv 0$ if $x<0$.  Note that we only examine violations of the lower pressure bounds $\pi^{\min}$ in the simulation, because by construction compressor control is bounded at a maximum discharge pressure.%

\subsubsection{Generator curtailment analysis} 
To avoid violations of the lower pressure limits, gas supply to gas-fired power plants may be curtailed, leading to a reduction in electric power output. This represents a shortfall in generation capacity, which must be covered by other electric generation resources in an ad-hoc manner, and which increases operational risk. To assess this risk, we quantify the impact of reduced gas supply in terms of generation capacity curtailment.

For each scenario $\omega(t)$, the desired gas \avz{consumptions of} the gas-fired power plants \avz{are} given by $$d_{\mathrm{desired}}(t)=q_{\mathrm{heat}}(\tilde{p}(t)),$$
where $\tilde{p}(t)$ is determined by the generation control policy \eqref{eq:gen_policy}. 
The expected amount of curtailments is characterized by solving an  adapted version of the deterministic optimal gas flow scheduling \eqref{eq:dogf}. This problem maximizes the gas \avz{consumptions $\mathbf{d}$ of} the generators, with the upper bound set equal to the desired withdrawals
$d^{\max}=d_{\mathrm{desired}}(t)$ and the compressor schedule $\alpha(t)$ fixed to the solution obtained from Formulation 0, 1 or 2, respectively. 
The optimized gas delivery, which may be below $d_{\mathrm{desired}}(t)$, is then translated back through the heat rate to an adapted generation schedule. This schedule is used to compute the generation capacity curtailment $\Delta p$,
\begin{equation}
    \Delta p=\frac{1}{N_h}\sum_{t\in\mathcal{T}}\sum_{j\in\mathcal{G}_{\textrm{gas}}} (\tilde{p}_j(t)-q_{\mathrm{heat}}^{-1}(\mathbf{d}_j(t))).
    \label{eq:gen_curtailment}
\end{equation}
Here, $N_h$ is the number of time steps per hour and $\mathcal{G}_{\textrm{gas}}$ represents the set of gas-fired generators. The units of the generation capacity curtailment is [MWh], which represents the total energy curtailment (across all gas-fired generators) across the entire day.

\section{Case Study: Computational Results}
\label{sec:results}
With the above case study set up, we compare the performance of our proposed method (Formulation 2) against the two other formulations that do not account for the impact of reserve activation on the \avz{gas pipeline network}, and assumes either predetermined reserve capacities (Formulation 0) or reserve capacities scheduled using the chance-constrained optimal power flow (Formulation 1). We focus our attention on constraint violations in the natural gas network, \avz{because} previous work has analyzed the performance of similar chance-constrained optimal power flow formulations \cite{bienstock2014chance, roald2013analytical, roald2016corrective}.

\subsection{Comparison of Cost and Schedules}
We first analyze the cost and the schedules for the three formulations. The objective function values, as well as the scheduled generation and reserve capacities from gas generators are shown in Table \ref{tab:cases}.
Note that the cost of energy used by the compressors is several orders of magnitude smaller than the cost of electricity generation. The generation and compression cost is almost the same for Formulation 0 and 1, while Formulation 2 has approximately 1\% higher cost. This modest increase in cost \avz{arises from} the extra constraints that are included to guarantee robustness. Note also that all three formulations schedule a similar amount of generation from the \avz{gas-fired} generators, while Formulation 0 allocates \avz{to them} a lower amount of reserve capacity. This may be due to the fact that Formulation 0 does not account for the impact of the reserves on power system constraints such as line flow limits. 

\begin{table}%
\centering
\caption{Scheduling  Solutions} 
\begin{tabular}{|c|c|c|c|c|c|c|}
\hline
\hline
     Formulation  & Power             & Compression & Nominal Gas      & Reserve Gas   \\
     (90\%  load level) & Cost $J_P$  & Cost \avz{$J_C$}  & Generation  $\mathbf{\generationsetpoint}$  & Capacity $\mathbf{r}$ \\
        &   [\$M]  & [\$] & [MW] &  [MW] \\
\hline
\hline
Formulation 0      &  4.88  & 401.5  & 17482.1 & 2291.0  \\
Formulation 1     & 4.89  &  401.4   & 17482.0 & 2683.0   \\
Formulation 2    &  4.94  &  425.5 & 17482.0 & 2683.0 \\
\hline
\end{tabular}
\label{tab:cases}
\end{table}

While the total amount of generation and reserves scheduled from gas-fired generators is similar for all formulations, the \avz{allocation is not homogeneous}. Relative to the other two formulations, Formulation 2 schedules less generation from G2 and more generation from G3. Furthermore, generator G2 also provides less reserves in Formulation 2, while G1 and G4 provide more. To support the higher gas withdrawals at node J24 (which supplies G3) and more variable withdrawals at node J19 (which supplies G4), compressor C4 has a higher compression ratio. This demonstrates how the needs of the electric grid can impact \avz{gas pipelines}. %

\subsection{Pressure Violations and Generator Capacity Curtailment}
\begin{figure*}[!th]
    \centering
    \includegraphics[scale=0.8]{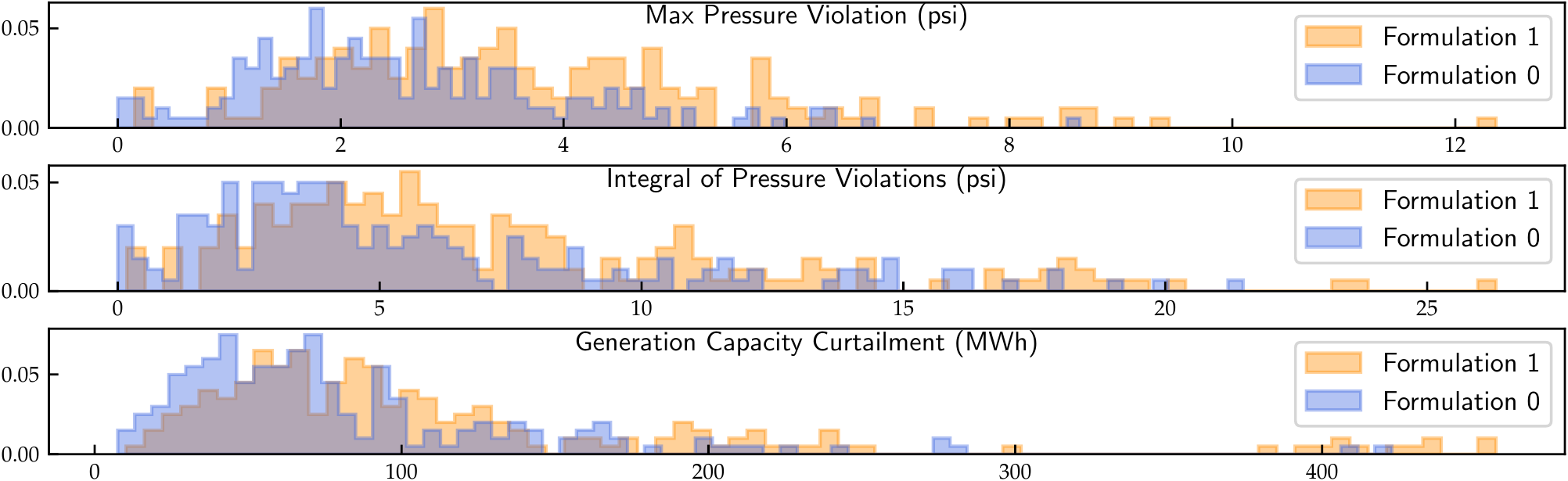}
    \caption{\textbf{Empirical probability distribution of the pressure violations and generator capacity curtailment.} Monte Carlo results that quantify how insufficient gas packed in the pipeline lead to curtailment of generation.  Top: Maximum violation of minimum pressure constraints as given by Eq. \eqref{eq:pviol_max}.  Center: $L_2$ norm of pressure violation over the simulation interval in psi-hours, given by Eq. \eqref{eq:pviol_norm}. Bottom: Expected total load shed (MWh) per day, computed from solving Problem \eqref{eq:dogf} for scenarios with pressure violations. Formulation 2 is not included, because it does not result in any violations or curtailments.}
    \label{fig:2violations-box-plot}
\end{figure*}

To assess the quality of solutions to Formulations 0, 1 and 2, we calculate the maximum pressure violation \eqref{eq:pviol_max}, the $L_2$ norm of the pressure violation \eqref{eq:pviol_norm}, and the corresponding generation capacity curtailment \eqref{eq:gen_curtailment} for each scenario. 
The average values of the pressure violations and curtailment across all scenarios \avz{are} shown in Table \ref{tab:violations}, while Figure \ref{fig:2violations-box-plot} shows the probability distribution derived from the Monte Carlo samples.

\begin{table}%
\centering
\caption{Statistics from Monte Carlo Simulation} 
\begin{tabular}{|c|c|c|c|c|c|c|}
\hline
\hline
     Formulation  & Mean Max. &   Mean Integrated  &  Mean Load \\
    (90\%  load level) & Pressure Viol. [psi] & Pressure Viol. [psi-hrs] &  Shed [MWh]\\
\hline
\hline
Formulation 0      &  2.61  & 5.69  &  81.9       \\
Formulation 1      & 3.67   & 7.32  & 115.9           \\
Formulation 2     & 0.00   & 0.00  &  0.00          \\
\hline
\end{tabular}
\label{tab:violations}
\vspace{-2ex}
\end{table}

From Table \ref{tab:violations}, we observe that our proposed joint scheduling and uncertainty management method (Formulation 2) does not lead to violations or generator capacity curtailments in any of the scenarios. This is as expected, because the proposed formulation guarantees gas \avz{pipeline} network feasibility for provision of gas for both energy and reserves. The other two formulations, which do not consider the \avz{linepack} required to \avz{guarantee} reserves, lead to \avz{minimum pressure constraint violations} and generator capacity curtailments. The maximum \avz{amount of} pressure violations is 2.6 psi (Formulation 0) and 3.7 psi (Formulation 1). 
Over the course of the day, the pressure violations result in an average curtailment in generation from gas-fired power plants of 81.9 MWh and 115.9 MWh per day for Formulations 0 and 1, respectively. While we do not explicitly model the impact of this shortfall in energy on the electric system, it clearly impacts the system operator's ability to maintain safe operations.

\begin{figure}%
    \centering
    \includegraphics[width=.75\linewidth]{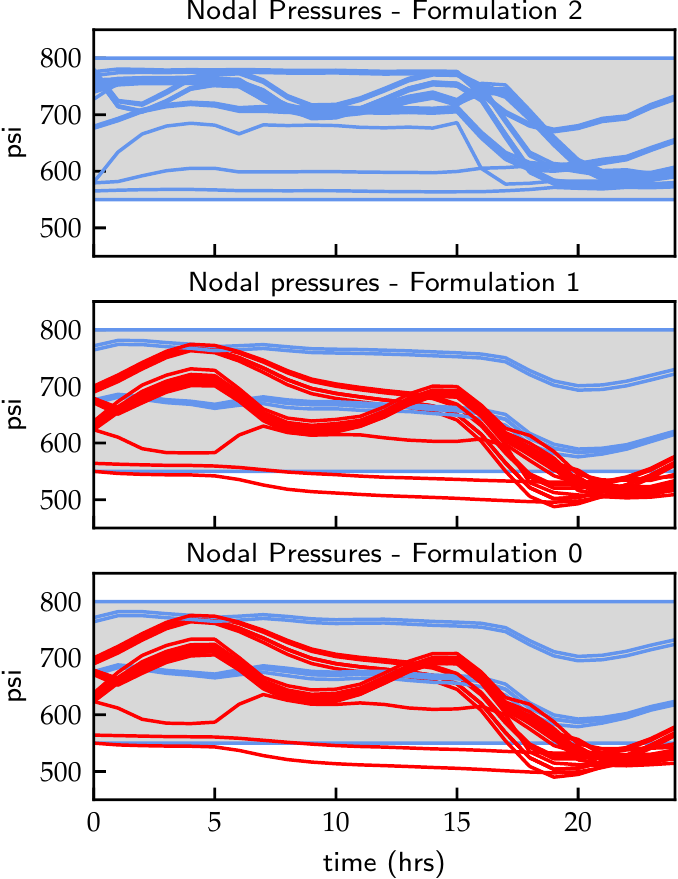}
    \caption{\textbf{Pressure trajectories for the maximum gas withdrawal scenario.} %
    We simulate the case with maximum gas withdrawal (corresponding to full activation of reserves) for the solutions corresponding to Formulation 2 (top), Formulation 1 (middle) and Formulation 0 (bottom). The grey area represent the range of feasible pressures. The lines represent the pressure trajectories at each node, including both feasible (blue) and infeasible (red) trajectories.}
    \label{fig:pressure-violations}
    \vspace{-2ex}
\end{figure}

While the above numbers give an indication of the average performance in terms of pressure violations, we analyze the scenario corresponding to the maximum gas withdrawals. For this scenario, we plot pressure trajectories for each node in the system in Figure \ref{fig:pressure-violations}. For the proposed method (Formulation 2), all pressure trajectories stay within the the limits (area shaded in grey). \avz{In Formulations 0 and 1, the pressure trajectories start} from the same feasible point, then rapidly \avz{lose} feasibility. Towards the end of the day, a majority of the nodes (with pressure trajectories marked in red) are below the lower pressure bounds. We observe that the pressure violations in this scenario are much more significant than the average violations, with some pressures dropping more than 50 psi below the limit.

\subsection{Robustness of the Joint Uncertainty Management}
\begin{figure}[!t]
    \centering
    \includegraphics[width=0.9\linewidth]{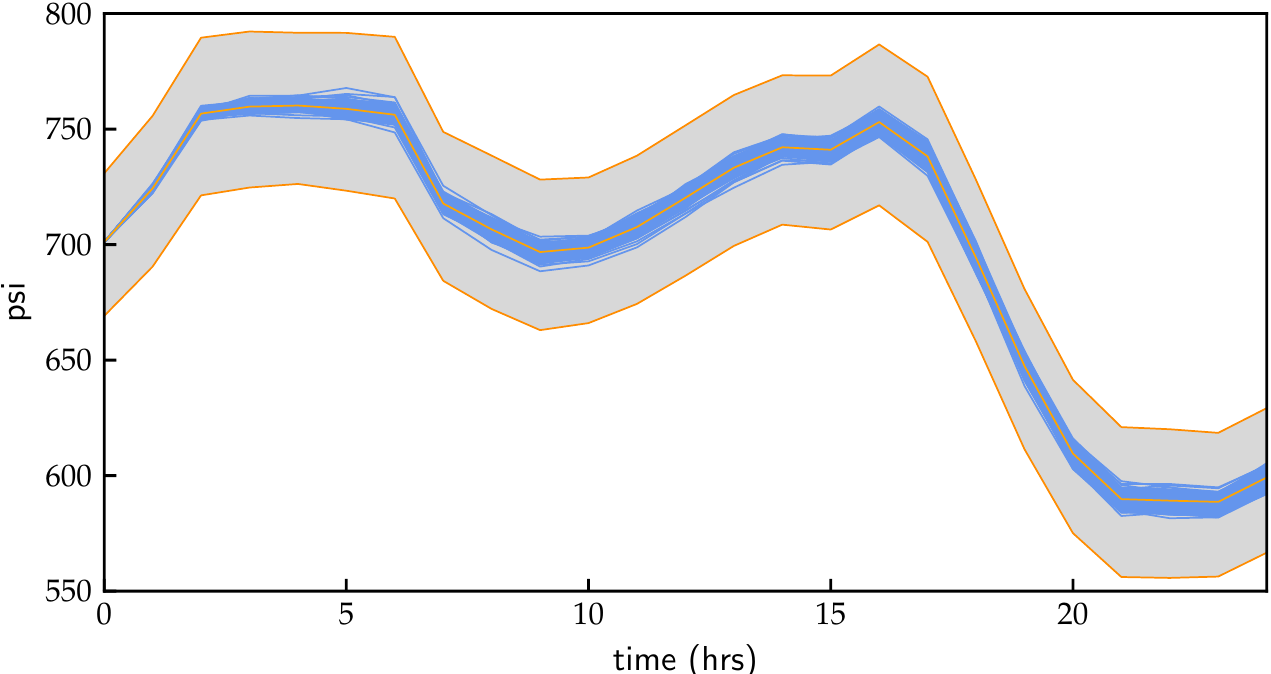}
    \caption{\textbf{Robust pressure trajectories for node 19.} We show the pressure trajectories for gas node 19 (power node 7). The yellow lines are the nominal, upper and lower pressure profiles obtained from Formulation 2, corresponding to the nominal, minimum and maximum withdrawal levels $\textbf{d}^{\textrm{nom}},\textbf{d}^{\min}$ and $\textbf{d}^{\max}$. The area between them is shaded in grey. The blue lines are pressure trajectories from the Monte Carlo simulation, which start at a nominal initial state, and diverge depending on reserve activation in the power system.}
    \label{fig:pressure-cone}
\end{figure}

To demonstrate some characteristics of the robust gas system formulation, we investigate the pressure at one particular node (gas node 19, which is connected to electric node 7). Figure \ref{fig:pressure-cone} shows the pressure trajectories corresponding to the nominal, maximum and minimum withdrawal cases from the joint scheduling and uncertainty management problem (plotted in yellow), as well as the pressure trajectories from the Monte Carlo simulations (plotted in blue). All the simulated trajectories start from the same state as the nominal pressure trajectory.
We observe that the Monte Carlo pressure trajectories initially follow the nominal trajectory, but then slowly spread out over a larger range. However, after approximately 12 hours, the spread seems to stabilize and not increase further. Furthermore, the simulated trajectories remain far away from the yellow lines representing the pressure trajectories of the minimum and maximum withdrawal cases. 

The behavior of the Monte Carlo trajectories can be explained by considering the distribution of the reserve activation. The deviations in the power loads $\omega(t)$ are modeled as a multivariate normal distribution, with no spatial or temporal correlation. This implies that the reserve activation, which is based on the total power imbalance $\Omega(t)$, and the corresponding gas consumptions are also normally distributed, with independent values at each time step. Over time, the gas withdrawals required for reserve activation hence \avz{average to} zero. 
Therefore, the solution mostly adjusts \emph{when} generators withdraw gas, as opposed to \emph{how much} gas is withdrawn in total (although there are of course differences). Consequently, the pressure trajectories do not deviate very far from the nominal. If we instead had a consistent activation of the upward or downward reserves, we expect a much larger spread in the observed pressure trajectories.

Finally, we would like to note that our formulation is not only robust to variations in the gas withdrawals, but also robust with respect to the initial conditions. Although we chose to start all Monte Carlo simulations from the system state corresponding to the nominal gas withdrawals, any initial condition with pressures in the range between (and including) the yellow lines would be robustly feasible to variations in the gas withdrawals. This can be useful for operational planning purposes, when the exact operating point at the beginning of the next day is not yet known.

\section{Conclusions}
\label{sec:conclusion}

This paper reviews the emerging trends in the energy sector that motivate integrated optimization and coordination mechanisms for gas pipelines and the electric grid, with a particular focus on managing intra-day uncertainty. One of the open challenges identified in the literature review is a method for joint optimization of generation, allocation of reserves, and gas pipeline flows under consideration of uncertainty \avz{and transient gas pipeline flows}. To tackle this challenge, we propose a framework that defines the interface between the electric and gas \avz{networks} as a nominal schedule, as well as upper and lower bounds on the gas \avz{consumed} by gas-fired power plants. Our proposed formulation builds on established approaches for \emph{chance-constrained optimal power flow} and new ideas for \emph{robust optimization of transient natural gas flow}. The proposed approach results in a numerically tractable optimization problem that achieves integrated scheduling with probabilistic and robust safety guarantees. 
In our case study, we benchmark our proposed framework against other joint optimization formulations in a case study with interacting test networks. Through Monte Carlo simulations, we verify the robust guarantees of our approach and demonstrate that such guarantees are essential to ensure \avz{operating reliability of both gas pipelines and power grids with high renewable contributions.}

\let\oldbibliography\thebibliography
\renewcommand{\thebibliography}[1]{%
  \oldbibliography{#1}%
  \setlength{\itemsep}{0pt}%
}

\bibliographystyle{unsrt}
\bibliography{references/references.bib}

\begin{IEEEbiography}[
{\includegraphics[width=1in,height=1.25in,clip,keepaspectratio]{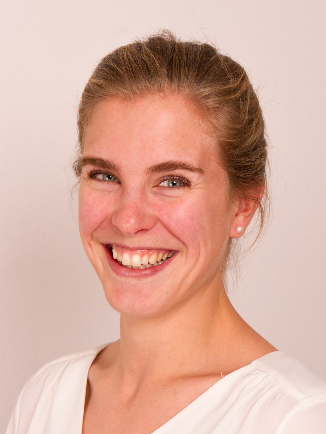}}]{Line Roald}  is an Assistant Professor and Grainger Institute Fellow in the Department of Electrical and Computer Engineering in University of Wisconsin--Madison.
She received her Ph.D. degree in Electrical Engineering (2016) and M.Sc. and B.Sc. degrees in Mechanical Engineering from ETH Zurich, Zurich, Switzerland. She was a postdoctoral research fellow with the Center of Non-Linear Studies at Los Alamos National Laboratory. Her research interests focus on modeling and optimization of energy systems, with a particular focus on managing uncertainty and risk from renewable energy variability and component failures.
\end{IEEEbiography}

\vspace*{-2em}
	
\begin{IEEEbiography}[
{\includegraphics[width=1in,height=1.25in,clip,keepaspectratio]{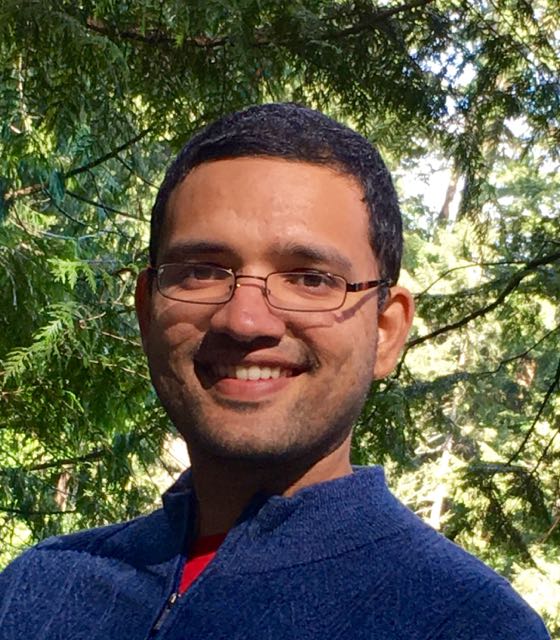}}
]{Kaarthik Sundar}
received the Ph.D. degree in mechanical engineering from Texas A\&M University, College Station, TX, USA, in 2016. He is currently a Research Scientist in the Information Systems and Modeling Division at Los Alamos National Laboratory, Los Alamos, NM, USA. His research interests include problems pertaining to vehicle routing, path planning, and control for unmanned/autonomous systems; numerical optimal control, estimation, and large-scale optimization problems in power and gas networks; combinatorial optimization; and global optimization for mixed-integer nonlinear programs.
\end{IEEEbiography}

\vspace*{-2em}

\begin{IEEEbiography}[
{\includegraphics[width=1in,height=1.25in,clip,keepaspectratio]{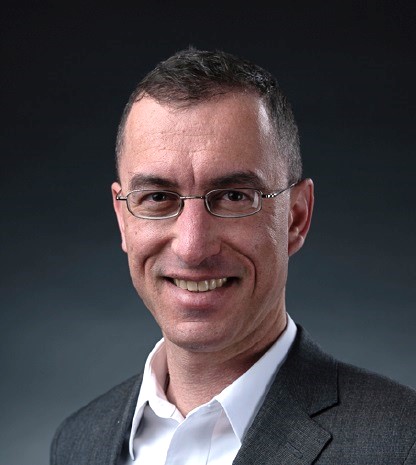}}
]{Anatoly Zlotnik (M10)} is a staff scientist in the Applied Mathematics and Plasma Physics group of the Theoretical Division at Los Alamos National Laboratory, where he was previously a postdoctoral associate at the Center for Nonlinear Studies.  Before joining LANL in 2014, he obtained a Ph.D. in systems science and mathematics from Washington University in St. Louis, Missouri, an M.S. in applied mathematics from the University of Nebraska – Lincoln, and B.S. and M.S. degrees in systems and control engineering from Case Western Reserve University in Cleveland, Ohio.  His research focus is on computational methods for optimal control of large-scale nonlinear dynamic systems.
\end{IEEEbiography}

\vspace*{-2em}

\begin{IEEEbiography}[
{\includegraphics[width=1in,height=1.25in,clip,keepaspectratio]{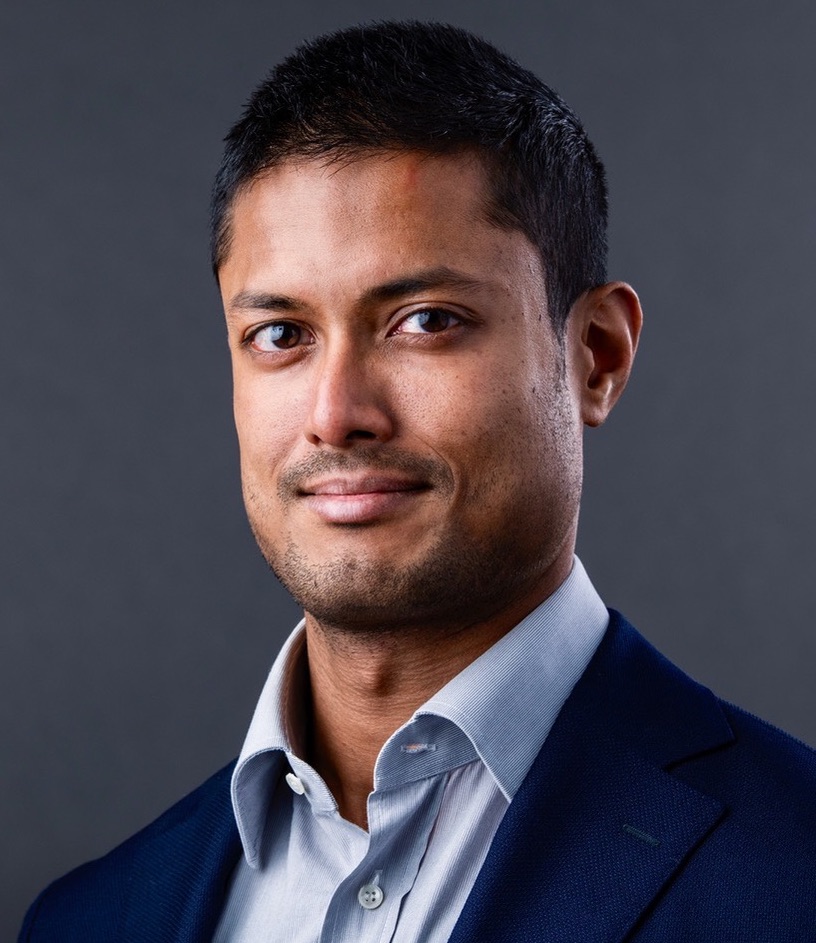}}
]{Sidhant Misra}
Sidhant Misra received the Ph.D. degree in electrical engineering and computer science from the Massachusetts Institute of Technology, Cambridge, MA, USA, in 2014. He is currently a Research Scientist in the theory division and the Advanced Network Science Initiative at the Los Alamos National Laboratory, Los Alamos, NM, USA. His research interests lie at the intersection of machine learning and optimization with applications in network structure identification, and planning and operations of energy networks.
\end{IEEEbiography}
	
\vspace*{-2em}

\begin{IEEEbiography}[
{\includegraphics[width=1in,height=1.25in,clip,keepaspectratio]{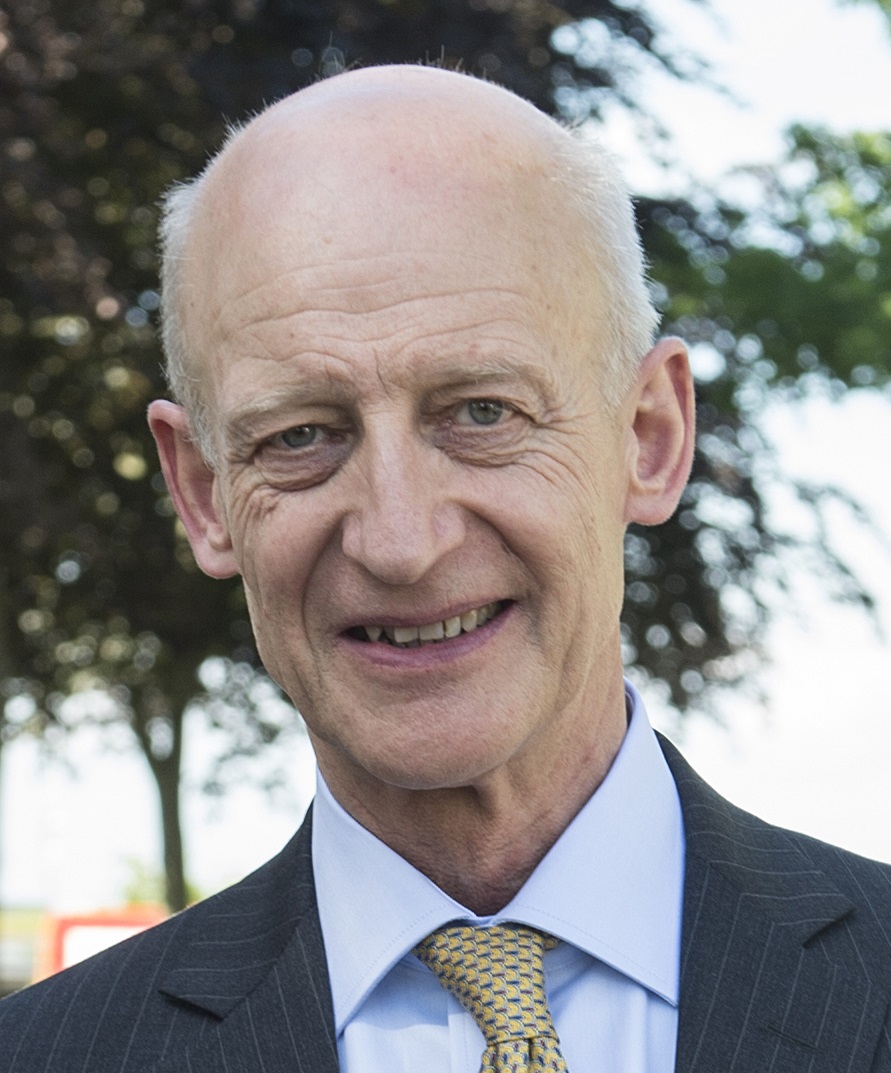}}]{G\"oran Andersson}  (M’86, SM’91, F’97) obtained his M.S. (1975) and Ph.D. (1980) degrees from the University of Lund, Sweden. In 1980 he joined ASEA’s, now ABB’s, HVDC division in Ludvika, Sweden, and in 1986 he was appointed professor in electric power systems at KTH (Royal Institute of Technology), Stockholm, Sweden. In 2000 - 2016 he was full professor in electric power systems at ETH Z\"urich (Swiss Federal Institute of Technology). His research interests include power system dynamics, control and operation, power markets, and future energy systems.
G\"oran Andersson is Fellow of the Royal Swedish Academy of Sciences, the Royal Swedish Academy of Engineering Sciences, the Swiss Academy of Engineering Sciences, and foreign member of the US National Academy of Engineering. He was the recipient of the 2007 IEEE PES Outstanding Power Educator Award, the 2010 George Montefiore International Award, and the 2016 IEEE PES Prabha S. Kundur Power System Dynamics and Control Award.
\end{IEEEbiography}

\vspace*{-2em}

\end{document}